\documentclass[aps,pra,preprint,groupedaddress,amsmath,amssymb]{revtex4}

\usepackage[bookmarks,bookmarksopen]{hyperref}
\usepackage{graphicx}
\usepackage{dcolumn}
\usepackage{bm}
\def\ref#1{$^{#1}$}
\newcommand{\aap}{A\&A}
\newcommand{\apjl}{ApJL}

\newcommand{\araa}{Ann. Rev. Astron. Astrophys.}
\newcommand{\mnras}{MNRAS}
\newcommand{\planss}{Planetary Space Science}
\begin{document}

\preprint{APS}
\preprint{Submitted to Int. Rev. Phys. Chem.}

%\title{A new theoretical study of the para--ortho-H$_2$ conversion due
%  to H collisions : Comparison with experiments}

\title{Ortho--para-H$_2$ conversion processes in astrophysical media}

\author{Fran\c{c}ois Lique} \email{francois.lique@univ-lehavre.fr}
\affiliation{LOMC - UMR 6294, CNRS-Universit\'e du Havre, 25 rue Philippe Lebon, BP 540, 76058, Le Havre, France}%
\author{Pascal Honvault} \affiliation{Laboratoire ICB, UMR 6303, CNRS-Universit\'e de Bourgogne, 21078 Dijon cedex, France and UFR Sciences et Techniques, Universit\'e de Franche-Comt\'e, 25030 Besan\c con Cedex, France}
\author{Alexandre Faure} 
\affiliation{UJF-Grenoble 1/CNRS, Institut de Plan\'etologie et d'Astrophysique de Grenoble (IPAG) UMR 5274, Grenoble F-38041, France}

\date{\today}% It is always \today, today,

\begin{abstract}

We report in this review recent fully-quantum time-independent
calculations of cross sections and rate constants for the gas phase
ortho-to-para conversion of H$_2$ by H and H$^+$. Such processes are of
crucial interest and importance in various astrophysical environments. The
investigated temperature ranges was 10$-$1500~K for H+H$_2$ and
10$-$100~K for H$^+$+H$_2$. Calculations were based on highly accurate
H$_3$ and H$_3^+$ global potential energy surfaces. Comparisons with
previous calculations and with available measurements are presented
and discussed. It is shown that the existence of a long-lived
intermediate complex H$_3^+$ in the (barrierless) H$^+$+H$_2$ reaction
give rise to a pronounced resonance structure and a statistical
behaviour, in contrast to H+H$_2$ which proceeds through a barrier of
$\sim 5000$~K. In the cold interstellar medium ($T\leq 100$~K), the
ortho-to-para conversion is thus driven by proton exchange while above
$\sim$300~K, the contribution of hydrogen atoms become significant or
even dominant. Astrophysical applications are briefly discussed by
comparing, in particular, the relative role of the conversion
processes in the gas phase ({\it via} H, H$^+$, H$_2$ and H$_3^+$) and
on the surface of dust particles. Perspectives concerning future
calculations at higher temperatures are outlined.

\end{abstract}

\pacs{Valid PACS appear here}% PACS, the Physics and Astronomy
\maketitle

\tableofcontents

\newpage

\section{\label{sec:level1}Introduction}

Hydrogen is the most abundant element in the Universe and molecular
hydrogen, H$_2$, is the dominant molecule in all astrophysical
environments, from the atmospheres of giant (exo)planets to external
galaxies.  Molecular hydrogen is also used in many industrial
applications, and is often discussed as a potential energy carrier for
the future.  In particular, low-temperature hydrogen plasmas are
relevant for many technological plasma applications including fusion.
In the interstellar medium (ISM), where stars and planets form, H$_2$
is produced efficiently {\it via} the recombination of hydrogen atoms
on the surface of sub-micron size dust grains \citep{watanabe08}. In
the early Universe, where dust grains were absent, the dominant source
of H$_2$ was the associative detachment between H and H$^-$ and, at
high density, the three body recombination H+H+H
\citep{galli13}. Molecular hydrogen is a major contributor to the
physics and chemistry of astrophysical media and, in particular, it
played a fundamental role in the cooling of the gas clouds that gave
birth to the very first stars.

Owing to its identical hydrogen nuclei (with nuclear spin 1/2), H$_2$
exists in ortho (o-H$_2$) and para (p-H$_2$) forms, also called
nuclear-spin isomers. In the electronic ground state,
the rotational levels of o-H$_2$ have odd values of the angular
momentum $j$ while the levels of p-H$_2$ have even $j$ values. The
internal energy of the newly formed H$_2$ molecules is expected to
depend on the specific formation mechanism but it is generally assumed
that H$_2$ is initially highly excited. The ortho-to-para ratio (OPR)
of nascent H$_2$ is therefore usually taken as its limiting (high
temperature) statistical value of 3, which is the ratio of the
degeneracies of the ortho ($I$=1) and para ($I=0$) nuclear spin
states. The OPR of H$_2$ formed on cold ($<$50~K) solid surfaces was
studied in several recent experiments and the measured values were
indeed found to be consistent with the high temperature limit of 3
(see \citep{fukutani13} and references therein). In other processes,
the conservation of the total nuclear spin plays a crucial role and,
for instance, in the dissociative recombination of H$_3^+$ with
electrons, the OPR of the product H$_2$ molecules depends on the OPR
of the reactant H$_3^+$ molecules (see \cite{pagani09} and references
therein).

In an isolated state, the (radiative) interconversion between the
ortho and para states of H$_2$ is extremely slow, the theoretical time
scale being $\sim 5\times 10^{20}$~s \cite{pachucki08}, i.e. greater
than the age of the Universe. Ortho-to-para conversion (OPC) (also referred as nuclear spin conversion) is also
forbidden in non-reactive inelastic collisions. As a result, the OPC
can only occur in the gas phase {\it via} ``spin exchange'' (reactive)
collisions and in the solid phase {\it via} interaction with surfaces
of magnetic or diamagnetic materials, including amorphous water ice
\cite{sugimoto11}. In both laboratory and space environments, the OPR
of H$_2$ is however not necessarily at thermal equilibrium because the
timescale of the OPC process can be significantly longer than the
timescale of the thermal evolution (and also than the thermalization
time within each modification). Thus, the H$_2$ OPR measurements
reported in the astronomical literature often show values out of
equilibrium with the environment temperature, e.g. in planetary
atmospheres \citep{huestis08} or in protostellar shocks
\citep{neufeld98,maret09} where the values typically vary from 1 to
3. The OPR value therefore provides a valuable probe of the thermal
history and lifetime of the observed astronomical object. In cold
interstellar clouds, where H$_2$ is invisible, the OPR is expected to
decrease slowly with temperature (from its initial value of 3) but to
remain higher than the thermal equilibrium value ($\sim 4\times
10^{-7}$ at 10~K) \citep{flower06}. Indirect evidences based on the
observation of deuterated molecules (see \cite{flower06,pagani13} and
references therein) and ammonia \citep{dislaire12,faure13} indeed
suggest OPR values of $\sim$10$^{-3}$. This has important consequences
because the OPR of H$_2$ not only affects the interstellar chemistry
but also the molecular excitation (see e.g. \cite{troscompt09}) and
therefore the cooling by molecular line emission.

In astronomical environments, the OPC rate on solid surfaces is highly
uncertain because it strongly depends on poorly known parameters such
as the electronic structure of the surface, the residence time of
H$_2$, the conversion efficiency, etc. \cite{lebourlot00}. In the gas
phase, the conversion can occur through spin-scrambling reactions
between H$_2$ and the most abundant hydrogenated species H, H$^+$,
H$_2$ and H$_3^+$. The relative contribution of each colliding partner
depends on its relative density and on the kinetic
temperature. Indeed, because the reactions of H$_2$ with H and H$_2$
have substantial barriers ($\sim$ 5000~K and $\sim$ 60 000~K,
respectively), the OPC in the cold ISM (i.e. $j=1\rightarrow 0$) is
driven by proton exchange with H$^+$ and H$_3^+$. In warmer regions
($\gtrsim 300$~K), however, reactive collisions with H become
efficient. Examples are shock-heated gas, photon dominated regions
(PDRs), supernova remnants, and the primordial gas. Finally, in
planetary atmospheres where the H$_2$ density is much larger than in
the ISM ($\gtrsim 10^{20}$~cm$^{-3}$), H$_2$+H$_2$ collisions can play
an important role despite very low rate coefficients
\citep{huestis08}.

In the present work, we review recent theoretical work on the OPC of
H$_2$ by hydrogen and proton exchange in the gas phase. These two
reactions, H+H$_2$ and H$^+$+H$_2$ (and their isotopic analogs), have
become prototypes for triatomic reactions and because of their
experimental and theoretical accessibility, they have have been
studied in much detail, allowing a very accurate understanding of the
dynamics. Surprisingly, however, the specific OPC process has not been
the object of many theoretical investigations in the past. Very
recently, it has been revisited using full dimensional quantum
time-independent approaches combined with high accuracy ab initio
potential energy surfaces (PES) for H$_3$ \cite{lique:12:H2}
and H$_3^+$
\cite{honvault:11,honvault11a,honvault12}. These
triatomic systems are indeed amenable to state-of-the-art
computational methods and various measurements have confirmed that
theory and experiment have now converged for such elementary
reactions, as discussed below. In contrast, only approximate
theoretical methods have been applied so far to the four- and
five-atom systems H$_2$+H$_2$ and H$_3^+$+H$_2$. For these important
processes, we refer the reader to other recent sources. Thus, for the
OPC of H$_2$ by H$_2$, theoretical and experimental results are
discussed in \cite{carmona07,huestis08}. For the OPC of H$_2$ by
H$_3^+$, the interested reader can find detailed information on the
various recent experimental and theoretical works in
\cite{grussie12,gomez12} and references therein.  Finally, for the
conversion on solid surface, we refer the reader to the very recent
review by Fukutani \& Sugimoto \cite{fukutani13}.

This review is organized as follows: Section II provides a description
of the theoretical methods that can be employed for triatomic reactive
systems. In Sec. III results are presented for the OPC of H$_2$ by
hydrogen and proton exchange. In Sec. IV, astrophysical applications
are discussed. Concluding remarks are drawn in Sec. V.

%According
%to available experimental data for amorphous and graphite surfaces,
%the OPC rate could reach 10$^{-14}$~s$^{-1}$ \cite{fukutami13}, which
%is comparable to the life time of a molecular cloud.

%The energy difference between the lowest states $j=0$ and $j=1$ is
%170~K, which is much larger than the temperature in the dense
%interstellar medium ($\sim$10~K). Thus, at 10~K and thermal
%equilibrium, the OPR of H$_2$ is close to zero ($\sim 3.7\times
%10^{-7}$). The OPR in the interstellar medium is however generally out
%of equilibrium and it may significantly affect the chemistry.

\section{Theoretical methods}

The computation of reactive rate constants usually takes place within
the Born-Oppenheimer approximation for the separation of electronic
and nuclear motions. Reactive cross sections are thus obtained by
solving the motion of the nuclei on an electronic PES, which is
independent of the masses and spins of the nuclei.

\subsection{Potential energy surfaces}

The PES must be accurate since the dynamical calculations are very
sensitive to the PES quality. The most accurate treatments to compute
PES are those based on modern methods of {\it ab initio} quantum
chemistry \cite{Werner:12}. The process of reactive collision between
a diatomic molecule and an atom requires
generally, for the PES calculation, the use of {\it ab initio} methods
based on configuration interaction (CI). CI
methods like multireference internally contracted
configuration-interaction (MRCI) method \cite{Werner:88,knowles:88},
which is currently the most accurate method, is generally used to
describe all geometries that could be explored by the nuclei during
the collisional process.

For light triatomic reactive systems like the H+H$_2$ and H$^+$+H$_2$
systems, Full Configuration Interaction (FCI) method
\cite{knowles:84,knowles:89} which provides numerically exact
solutions (within the atomic basis set) can also be used. This
approach is very CPU consuming but can provide a very accurate
description of the correlation energy as long as large atomic basis
set are used.

For heavier systems or with many degrees of freedom,
Multiconfigurational second-order perturbation theory (CASPT2)
methods \cite{Andersson:92,Celani:00} can alternatively be used in
order to keep reasonable CPU time without altering too much the
accuracy of the PES.

Finally, for reactive systems whose wave function can be reasonably
well described by a single determinant wave function, coupled-cluster
theory \cite{shavitt2009} can also provide an accurate estimate of the
PES. However, reactive systems are rarely monoconfigurational systems
and such approach cannot generally be used.
 
In the methods described above, the quality of the result is also
determined by the choice of atomic orbitals to describe the molecular
orbitals and the electronic configuration. The chosen atomic orbitals
basis set, from which are built the molecular orbitals, must be large
enough to correctly represent the correlation energy and not too
extended so that the computation time remains acceptable.  The
augmented correlation-consistent valence quadruple zeta (aug-cc-pVQZ)
or quintuple zeta (aug-cc-pV5Z) basis sets of Dunning and
co-corkers~\cite{kendall:92,woon:93,woon:94} are usually well adapted
to interaction PES calculations.  In order to get accurate results
with respect to the atomic basis set, the energies obtained with the
incomplete basis set can be extrapolated to the Complete Basis Set
(CBS) limit, employing for example the mixed exponential and Gaussian
formula \citep{peterson:94,feller:00}:
$E_X=E_{CBS}+Ae^{-(X-1)}+Be^{-(X-1)^2}$, where $X$ denotes the size of
the smaller basis and $E_{CBS}$, $A$ and $B$ are adjustable fitting
parameters.

The standard quantum chemistry methods described above are implemented
in several widely used numerical codes (MOLPRO \cite{molpro},
GAUSSIAN \cite{g09} or MOLCAS \cite{molcas})

Then, an additional important step related to the interaction PES
determination is the building of its analytic representation in order
to adequately perform the dynamical calculations.  Great care should
be taken in order to maintain the accuracy of the {\it ab initio} PES
in its analytical representation through elaborate fitting
techniques. Fitting methods such as the Reproducing Kernel Hilbert
Space (RKHS) method \cite{ho:96} or the double many-body expansion
(DMBE) \cite{varandas:2007} theory are generally used to obtain the
analytic representations.

Generally, the long range values derived from the {\it ab initio}
calculations are not very accurate as they result from the (small)
difference of two large numbers. One should keep in mind the
importance of carefully extending the analytical PES values to the
long range part which can be more precisely derived from perturbation
calculations as a $1/R $ expansion, where $R$ represents the distance
between the centers of charge of the two interacting systems. The long
range part is then described by the electrostatic, induction and
dispersion terms contributing to the total interaction energy of the
complex with the proper angular and radial dependences
\cite{Stone:96}. Small unphysical irregularities of magnitude $\simeq$
1 cm$^{-1}$ at long range can significantly affect the dynamical
calculations. These effects are even more crucial in the field of cold
and ultra-cold collisions.

The H$_3$ system considered in the present review can be viewed as a
prototype of small polyatomic systems.  The corresponding PES has been
extensively studied. The first fully {\it ab initio} PES was published
in 1978. Truhlar \& Horowitz \cite{Truhlar:78} made an accurate
least-squares fit to Liu and SiegbahnÕs calculations \cite{Siegbahn:78}
of the PES for the H+H$_2$ reaction. Approximately a decade later,
two refined versions of the H$_3$ PES, called DMBE \cite{Varandas:87}
and BKMP \cite{Boothroyd:91} have been published. Finally, the most
recent calculations by Boothroyd et al. \cite{Boothroyd:96} and Mielke
et al. \cite{Mielke02} have been widely used both for H+H$_2$
inelastic and reactive collisions, including the isotopic
variants. These two PESs generally lead to very good agreement with
available experimental data. The {\it ab initio} H$_3$ PES of Mielke
et al. \cite{Mielke02} that will be used for the results reviewed in
this work was calculated at the Full Configuration Interaction (FCI)
level using a complete basis set extrapolation. It is probably
one of the most accurate PES available for a chemical reactions. On
this PES, the reaction proceeds through a large barrier of $\simeq$ 0.4
eV. The van der Waals well associated to the H--H$_2$ complex in the entrance channel is about 20 cm$^{-1}$

About the H$_3^+$ system also considered in the present review,
several global PES covering the whole
configuration space are nowadays available in the literature. The
first global {\it ab initio} PES of the ground electronic state of
H$_3^+$ has been published in 2000 and it was based on full
configuration interaction (FCI) calculations with a high level basis
set \cite{aguado2000}. Global PESs of the three lowest electronic
singlet states of H$_3^+$ have been built from the
diatomics-in-molecule approach \cite{takayanagi2000,kamisaka2002} or
using the DMBE method \cite{viegas2007}.

\begin{figure}
\includegraphics[width=9.cm]{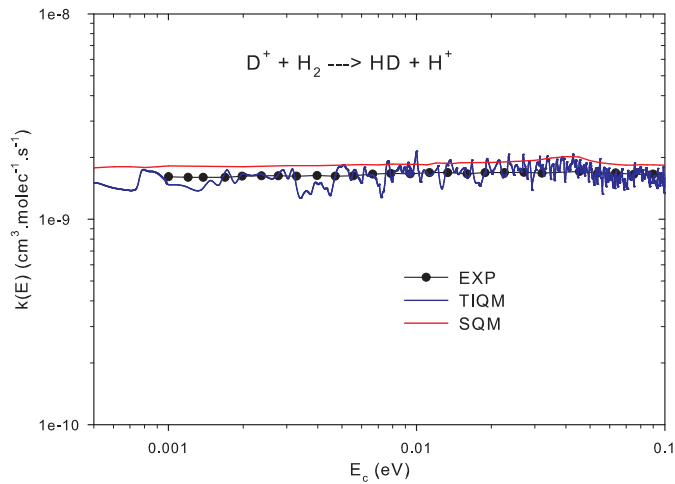}
\caption{Rate coefficients for the D$^+$ + H$_2(v=0,j=0)$ $\rightarrow$ HD + H$^+$ reaction. TIQM results (blue line) are compared with the Statistical Quantum Mechanical (SQM) prediction (red line) and the measurements. See the following reference for more details. Reprinted with permission from Ref. \cite{honvault13b}. Copyright 2013 American Chemical Society.}
\label{fig1}
\end{figure}

Recently, an updated version \cite{velilla2008} of the {\it ab initio}
PES of Aguado et al. \cite{aguado2000} has been published. In the
following section, the results on H$^+$+H$_2$ have been obtained on
this PES. We therefore recall the main features of this high quality
{\it ab initio} PES. Compared to the previous version published in
2000, new refinements have been indeed added in the present PES, such
as more {\it ab initio} points and the inclusion of a functional form of the
long-range electrostatic interaction in the analytical
representation. This last point is crucial to compute accurate
scattering attributes at low temperature.  In addition, the PES is
invariant under all permutations of the nuclei and presents a deep
well (4.6 eV relative to the H$^+$+H$_2$ asymptote) and no barrier
in the entrance channel. All approaches are therefore possible and
both abstraction and insertion mechanisms can occur.
In 2013, quantum mechanical and approximate dynamical calculations have been 
performed on this PES for the study of the D$^+$ + H$_2$ reaction \cite{honvault13a,honvault13b}.
The comparison between the quantum-mechanical rate constant and the 
measurements showed an excellent agreement as illustrated in Fig.~\ref{fig1}.
The very good accord between experiment and theory therefore validates the 
fact that this PES is sufficiently accurate to correctly describe the dynamics of such a reaction.

\subsection{Scattering Calculations}

Once the PES has been determined, the reactive cross sections and rate
constants are derived from the solution of the nuclear Schr{\"
  o}dinger equation within a given PES.  Several approaches can be
used to determine the reactive cross sections and rate constants. The
most widely used approach for reaction dealing with the H$_3$ and
H$_3^+$ systems are quantum time dependent and time independent
methods. Indeed, the large ro-vibrational energy level spacings of
H$_2$ make this system well suited for quantum scattering calculations.

Nevertheless, it is well known that the wave packet method may not be
very accurate at low collision energies that characterize the ISM
because of difficulties in damping outgoing waves with long de Broglie
wavelengths. A quantum time independent treatment seems to be the
most reasonable and accurate choice for such studies.

The computation of reactive cross sections is generally performed
by using methods based on a time-independent quantum formalism, called
`close-coupling" (CC). The quantal formalism has been introduced by
Arthurs and Dalgarno \cite{arthurs:60} for inelastic collisions and
then extended to reactive collisions by Schatz \&
Kuppermann \cite{Schatz:76}.
 
The three-dimensional time independent quantum mechanical reactive
scattering calculations usually employ a method based on body-frame
democratic hyperspherical coordinates \cite{Kupperman:80} to represent
the nuclear wave function.
In the present work, the ABC code \cite{Manolopoulos:00} and an alternative hyperspherical method \cite{Honvault:04} 
have been used to study respectively the dynamics of H+H$_2$ and that of the H$^+$+H$_2$ reaction.

An additional problem that is encountered when dealing with the H$_3$
and H$_3^+$ systems is that these are composed by three identical
nuclei that are indistinguishable. Contrary to what is usually the
case for other reactive systems, the entrance and exit channel cannot
be distinguished. H$_3$ and H$_3^+$ systems therefore belong to the
$P_3$ nuclear permutation symmetry group and the nuclear wave
function, nuclear spin excluded, belong to the $\Gamma=$ A$_2$ or E
irreducible representation. The specific symmetry properties of the
nuclear wave function can be incorporated in a simple manner, through
an appropriate choice of the hyperspherical harmonics that are built
from products of simple analytical functions \cite{Wu:91}.

When the three-particule system under considerations has $P_3$
permutation symmetry, the nuclear spin of the identical particles must
be taken into account. The total nuclear wave function can be
written as a direct product of the spatial wave function which
satisfies the spin-independent Schr\"odinger equation and a nuclear
spin wave function. For bosons, the total nuclear wave function is
symmetric, whereas, for fermions, it is antisymmetric with respect to
interchange of the identical nuclei. Accordingly, for the H+H$_2(vj)
\rightarrow$ H+H$_2(v',j')$ and H$^+$+H$_2(v,j) \rightarrow$ H$^+$+H$_2(v'j')$ 
reactions, the physically observable integral cross sections, $\sigma _{vj \to v'j' }$ 
that obey the proper spin
statistics, can be derived by weighting the calculated
integral cross sections \cite{Miller:69}:

$$ \left\{\begin{array}{lllll}
\sigma^{E}_ {vj \to v'j' } & \mbox{$j$ and $j'$ even ({\it para} $\rightarrow$ {\it para})} \\
{{2}\over{3}}\sigma^{A_2} _{vj \to v'j' } + {{1}\over{3}}\sigma^{E}_{vj \to v'j' } & \mbox{$j$ and $j'$ odd ({\it ortho} $\rightarrow$ {\it ortho})} \\
{{1}\over{3}}\sigma^{E} _{vj \to v'j' } & \mbox{$j$ odd, $j'$ even ({\it ortho} $\rightarrow$ {\it para})} \\
\sigma^{E} _{vj \to v'j' } & \mbox{$j$ even, $j'$ odd ({\it para} $\rightarrow$ {\it ortho})}
\end{array}
\right.
$$

It is worth mentioning that, if a full quantum method is used to treat
the dynamics of three identical nuclei, as in the present work, the
reactive and inelastic scattering processes cannot be distinguished.

Alternatively, cross sections may be calculated from wavefunctions
which treat the protons as distinguishable by appropriately adding 
scattering amplitudes for inelastic (e.g., $A
+ BC \to A + BC$) and reactive (e. g., $A + BC \to AB + C$ or $AC +
B$) processes obtained from such wavefunctions \cite{Truhlar:76}. In a
full description of the collision, the states in the H+H$_2$ or
H$^+$+H$_2$ arrangement are described by the quantum numbers $j$ and
$k$ (the rotational angular momentum of the H$_2$ molecule and its
projection along the reactant Jacobi vector), and $v$ (the vibrational
quantum number of the H$_2$ molecule). The integral cross section for
collision of H or H$^+$ with H$_2$($v,j$) to give H+H$_2$($v',j'$) or
H$^+$+H$_2$($v',j'$), summed over final projection quantum numbers and
averaged over initial projection quantum numbers, is given by:
	
\begin{widetext}

\begin{eqnarray}
\label{equ:cross_section}
 \sigma _{vj \to v'j' }(E_{c} ) =  \frac{\pi }{ {k_{vj }^{2} (2j + 1)}}
\sum \limits_{J kk'} (2J+1){\left| {S^{J } (E,vjk \to vj'k' )} \right|} ^{2} \nonumber \\
\end{eqnarray}
where $k_{vj}$ denotes the initial wavevector and where the $S$-matrix for the H+H$_2$ or H$^+$+H$_2$ reaction are given by :

{ \footnotesize 
\begin{eqnarray}
\label{Smat}
|S^J(E,vjk \to v'j'k')|^2 = 
\begin{array}{l}
 |S_n^J(E,vjk \to v'j'k')-S_r^J(E,vjk \to v'j'k')|^2,\\
 |S_n^J(E,vjk \to v'j'k')+S_r^J(E,vjk \to v'j'k')|^2 + 2|S_r^J(E,vjk \to v'j'k')|^2,  \\
3 |S_r^J(E,vjk \to v'j'k')|^2, \\
|S_r^J(E,vjk \to v'j'k')|^2, \\
\end{array}
\begin{array}{l}
j,j' even \\
j,j' odd \\
j \ even, j' odd \\
j \ odd, j' even \\
\end{array}
\end{eqnarray}
}

$S_n^J$ and $S_r^J$ are the nonreactive and reactive $S$-matrix
elements, respectively.
\end{widetext}

The summation in Eq. \ref{equ:cross_section} extends over all values
of the total angular momentum $J$ which contribute to the
reactive or inelastic process. The
scattering calculations are carried out on a grid of values of the
total energy $E$.  The relevant independent variable for the cross
sections is, however, the collision energy $E_c$, which is the initial
translational energy.  The two are related by
\begin{equation}
\label{equ:Ecol}
E_{tot}=E_c+\varepsilon_{vj}.
\end{equation}

where $\varepsilon_{vj}$ is the ro-vibrational energy of the H$_2$ reactant.

From the calculated cross sections $\sigma_{vj \to v'j'} (E_{c})$, one can obtain the corresponding thermal rate coefficients at temperature $T$ by performing a Maxwell-Boltzmann average over the collision energy ($E_c$):
\begin{eqnarray}
\label{thermal_average}
k_{vj \to v'j'}(T) & = & \left(\frac{8}{\pi\mu k_B^3 T^3}\right)^{\frac{1}{2}}  \nonumber\\
&  & \times  \int_{0}^{\infty} \sigma_{vj \to v'j'}(E_{c})\, E_{c}\, e^{\frac{-E_{c}}{k_B T}} dE_{c}
\end{eqnarray} 
where $\mu$ is the reduced mass of the reactive system and $k_B$ is the Boltzmann constant.

Note that the quantum mechanical study of the H$^+$+H$_2$ reaction
is more difficult than that of abstraction reactions, such as the H+H$_2$ reaction 
also presented in this review, for two principal
reasons. As already mentioned in the previous Section the
corresponding PES has a deep well and thus many states have to be
taken into account in the close-coupling equations. Moreover, in a
such complex forming reaction, symmetric top configurations (where the
Coriolis coupling is large) are energetically accessible. 
All (or nearly all) the allowed $\Omega$ components (where $\Omega$ is the projection of the total angular momentum $J$ on the axis of least inertia) have to be taken into account in the close-coupling expansion states, in order to obtain accurate cross sections. The maximum value of $J$, $J_{max}$, depends on the maximum of the collision energy employed in the scattering calculations. For the complex-forming H$^+$ + H$_2$ reaction detailed below, the maximum value of the collision energy is 0.1~eV. In this case, $J_{max}=35$ and $\Omega$ varying from 0 to 25 are enough for convergence of integral cross sections. By comparison, converged integral reactive cross sections have been obtained for the H + H$_2$ reaction up to 1~eV by using $J_{max}=75$ and restricting the projection quantum number of $J$ to 6 only.
Quantum mechanical calculations using the
hyperspherical method described above are therefore very consuming in
CPU time and in memory.  However, in contrast with the isotopic
variant D$^+$+H$_2$ \cite{honvault13a,gonzalez13,honvault13b} where
the same PES is used, the H$^+$+H$_2$ system offers a great
advantage, the three identical nuclei which noticeably reduce the CPU
time. The ABC code and the post-symmetrisation given by Eq.~2 have been used to study the H + H$_2$ reaction, while the other TIQM approach has been used to study the H$^+$ + H$_2$ reaction by considering the three nuclei as undistinguishable.
Recently, we checked the two methods gave similar results on the H + H$_2$ reaction \cite{lique:12:H2}. We showed the equivalence between results obtained by the post-symmetrisation of the S-matrix elements computed using the ABC code and results obtained from S-matrix elements with the correct exchange symmetry (using the other TIQM approach).

We should also mentioned that calculations of the reactive rate
constants can be done using quasi-classical trajectory (QCT) method
\cite{Bonnet:04,Mandy:09}. The QCT method combines the use of
classical mechanics, to treat the scattering process, but the
quantization of the reactants is taken into account. Quantization is
simulated by means of a `binning" procedure, which involves allocating
the final states to discrete values of the corresponding quantum
numbers. However, the QCT method is only valid as long as the
classical mechanics that underpins it. For low temperatures collisions
(as those found in the ISM), the QCT approach seems to be not very appropriate
because the inability of QCT treatments to conserve the vibrational
zero-point energy can render this method unreliable near reaction
thresholds.

Accurate statistical methods based on a pure quantum mechanical formalism, 
such as the statistical quantum mechanical (SQM) method \cite{RHM:CPL01,RGM:JCP03},
can also be used for the study of complex forming reactions such as the H$^+$+H$_2$ reaction.
An example of such calculations is given in the next section.

\section{Ortho--para-H$_2$ conversion processes}

\subsection{Ortho--para-H$_2$ conversion by hydrogen exchange}

Surprisingly, despite its crucial importance for the physical
chemistry of early Universe and of the ISM but also for the physical
chemistry of hydrogen plasmas, the OPC process of H$_2$ due to H
collisions has not been the object of many theoretical or experimental investigations.
Indeed, in most of the rotationally inelastic excitation studies of
H$_2$ by H, the rigid rotor approximation was used and therefore, the
reactive channels were neglected \cite{Flower97,Wrathmall06}. The goal
of these studies was the calculation of rotationally inelastic rate
coefficients for astrophysical applications at low to moderate
temperatures (below 1000 K). Rigid rotor approach could be justified
since the reaction between H and H$_2$ is inhibited by a large barrier
($\simeq$ 5000 K), and the reactive rate constants corresponding to
OPC of H$_2$ are expected to be very small at these
temperatures.

The OPC process of H$_2$ by hydrogen exchange has been
already studied by Truhlar \cite{Truhlar:76}, Mandy \&
Martin \cite{Mandy:92} and by Sun \& Dalgarno \cite{Sun:94}. The most
recent work of Sun \& Dalgarno \cite{Sun:94}, despite relatively
accurate, was restricted to p-H$_2(j=0)$ and o-H$_2(j=1)$ and was then
too limited for astrophysical applications of, for example, early
Universe. Experimentally, the para--ortho conversion process of H$_2$ was
measured in the temperature range 300-444~K almost 50 years ago by
Schulz \& Le Roy \cite{Schulz:65} and to the best of our knowledge, no
measurement have been done since then.

The H$_3$ electronic ground state, on which the H + H$_2$ dynamical calculations
are performed conically intersects the first excited state
\cite{Bouakline:10}.  The crossing occurs at an energy of 2.7 eV with
respect to the bottom of the H$_2$ ground electronic states well.  As
a consequence, there is a sign change of the electronic wavefunction
as one follows a closed path in nuclear configuration space around the
line of the conical intersection. This sign change is usually referred
as the geometric phase effect which can be globally taken into account
at low energies (E $<$ 15000~cm$^{-1}$) by reversing the sign of the
$S_r^J$ \cite{Mead:79,Lepetit:90} matrices using the theoretical
approach which treat the protons as distinguishable. The geometric
phase effect has been discussed in detail in reviews of Aoiz et
al. \cite{Aoiz05} and Bouakline et al. \cite{Bouakline:10}

Apart from the geometric phase effects, non-adiabatic transitions in
the region of proximity of the two electronic surfaces are also
possible (at high collision energies). However, for these systems, it
has been found that non-adiabatic transitions are unlikely to occur
\cite{Bouakline:10}. Then, H + H$_2$ reaction can be safely studied
within the Born-Oppenheimer approximation contrarily to what has been
found for others simple reactions such as F + H$_2$ \cite{Yang2007,Lique2011} or
Cl + H$_2$ \cite{Wang:08}.

Recently, we investigated the OPC process of H$_2$ by
hydrogen exchange \cite{lique:12:H2} (hereafter Paper I) within the
Born-Oppenheimer approximation. In our investigation of the scattering
dynamics, we used the H$_3$ global potential energy surface (PES) of
Mielke {\it et al.} \cite{Mielke02} that was proved to be very
successful in the calculation of H+H$_2$ thermal rate coefficients
\cite{Mielke03}. We used a pure quantum time independent approach in
order to get a very accurate modeling of this crucial process for the
astrophysical modeling of ``hot'' environment and of early Universe. In these calculations, we
considered results only for H$_2$ molecules in their ground vibrational states (despite excited vibrational levels were included in the calculations) and we
considered transitions between rotational states up to $j=10$.

First of all, in Paper I, we checked that, at low and intermediate
collisional energies, the geometric phase effect can be
neglected. Nevertheless, all results were obtained including the
geometric phase effect (by reversing the sign of the $S_r^J$).

\begin{figure}
\includegraphics[width=7. cm]{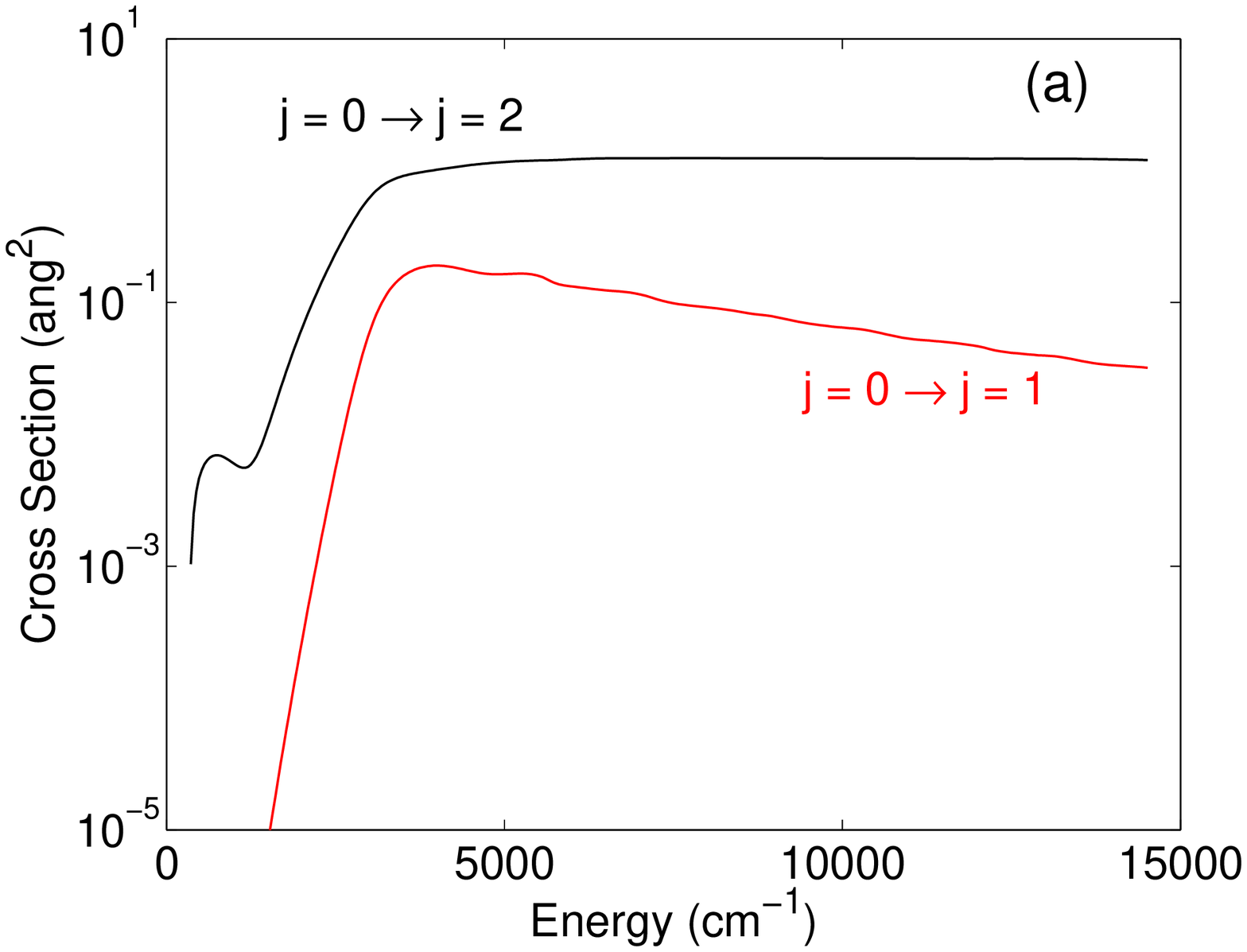}
\includegraphics[width=7. cm]{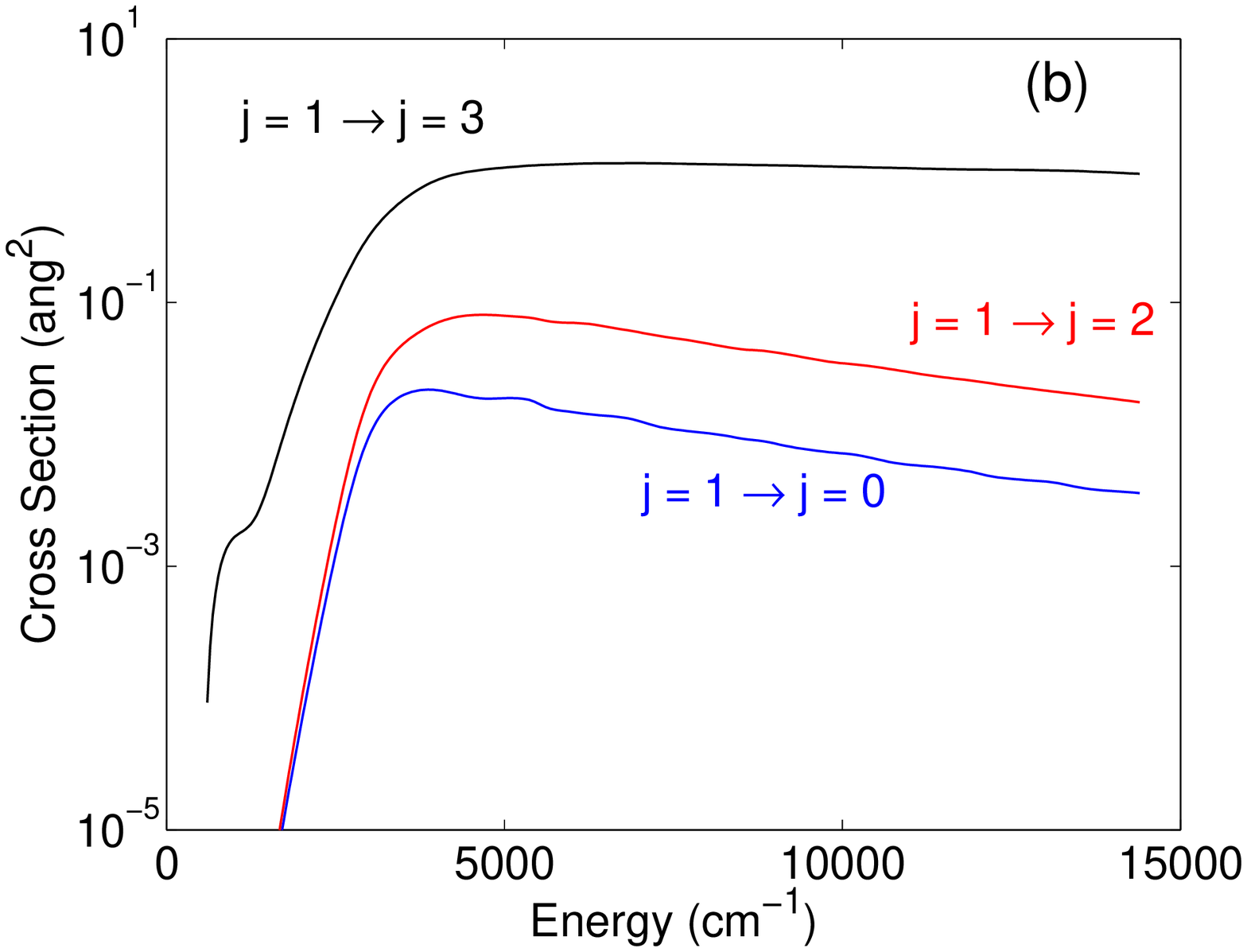}
\caption{Collision energy dependence of the integral cross section for
  the rotational excitation of p-H$_2(j=0)$ (a) and
  o-H$_2(j=1)$ (b) by H. Reprinted with permission from Ref.  \cite{lique:12:H2}. Copyright 2012 American Institute of Physics.}
\label{fig2}
\end{figure}

In Paper I, we have computed the collisional energy dependence of the
inelastic and reactive cross sections (considering the particle as
distinguishable) and we get the cross sections for the rotational
excitation of H$_2$ by H using the post-symmetrisation described
above.  Figure~\ref{fig2} displays the energy dependence of the
calculated integral cross sections for rotational (de-)excitation of
p- and o-H$_2$  by H, respectively.

As already discussed in Paper I, at low collisional energies, the
magnitude of the p--o-H$_2$ and o--p-H$_2$ cross sections is small and
in any case, these cross sections are much smaller than those for
transitions conserving the nuclear spin ($\Delta j=2$). OPC process of
H$_2$ is then relatively negligible at low energies compared to the
pure rotational excitation process.  It is interesting to note that,
at the opposite of what has been found for the same H$_2$ transition
due to H$^+$ collisions (see below), the de-excitation
H + o-H$_2(j=1)$ $\to$ H + p-H$_2(j=0)$ cross sections is negligibly
small at low collisional energies even if it is the only energetically
possible process. These results were expected and could have been
anticipated. Indeed, as the reaction proceeds by tunneling effect, the
magnitude of the cross sections is negligible at these low energies.

However, the magnitude of the cross sections increases rapidly with
increasing collisional energies and at collisional energy $\sim$
2000--3000~cm$^{-1}$, the spin conversion process becomes only one order
of magnitude smaller than the spin conserving process, showing the
clear competition between inelastic and reactive processes.

Taking into account the relatively large abundance of hydrogen atom in
diffuse ISM or in early Universe, the OPC process of H$_2$ by
hydrogen exchange remains a major process in the OPC
process of H$_2$ as soon as these collisional energies are encountered. The
temperature variation of the corresponding rate coefficients have 
been computed from the inelastic cross sections (Paper I).

\begin{figure}
\includegraphics[width=7. cm]{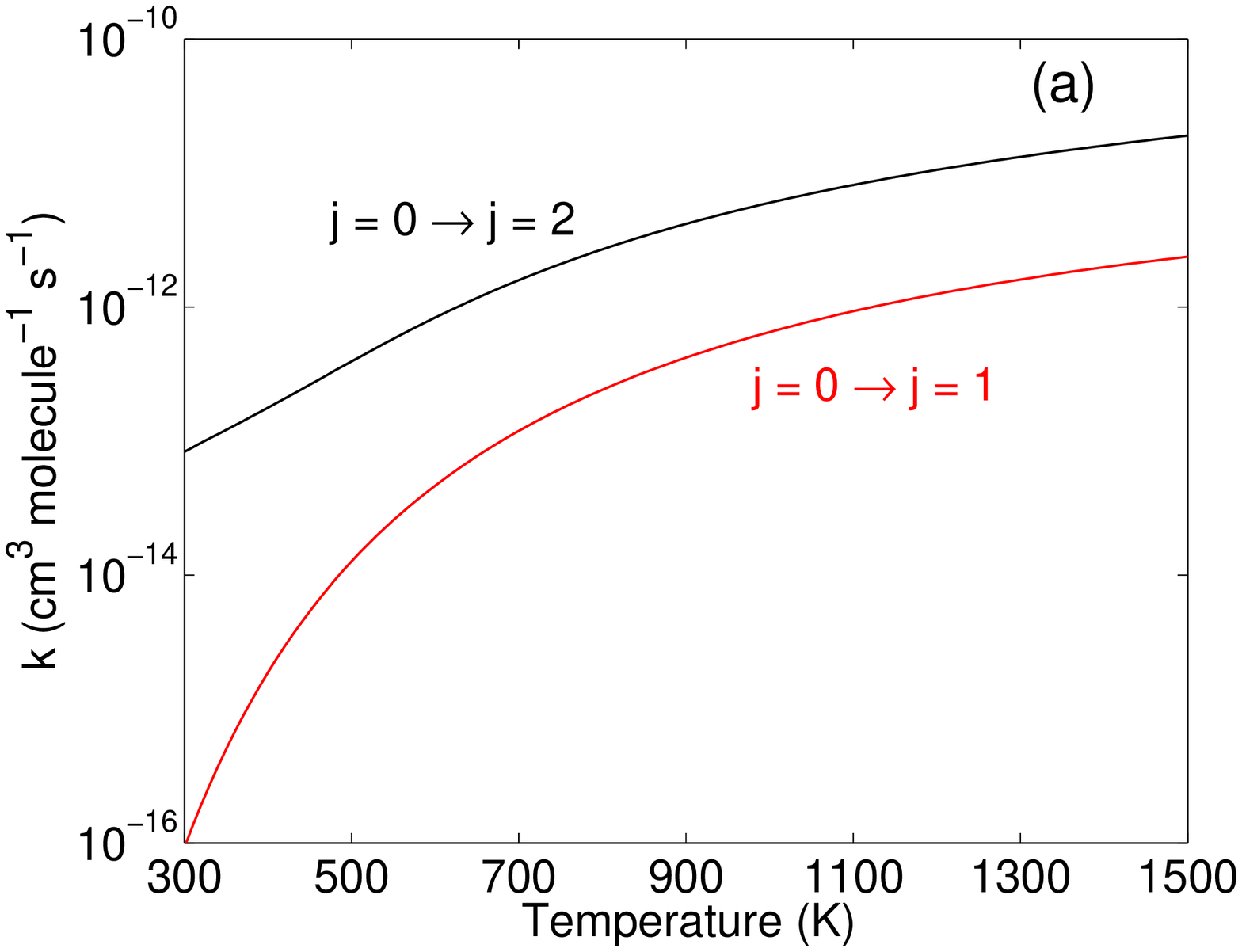}
\includegraphics[width=7. cm]{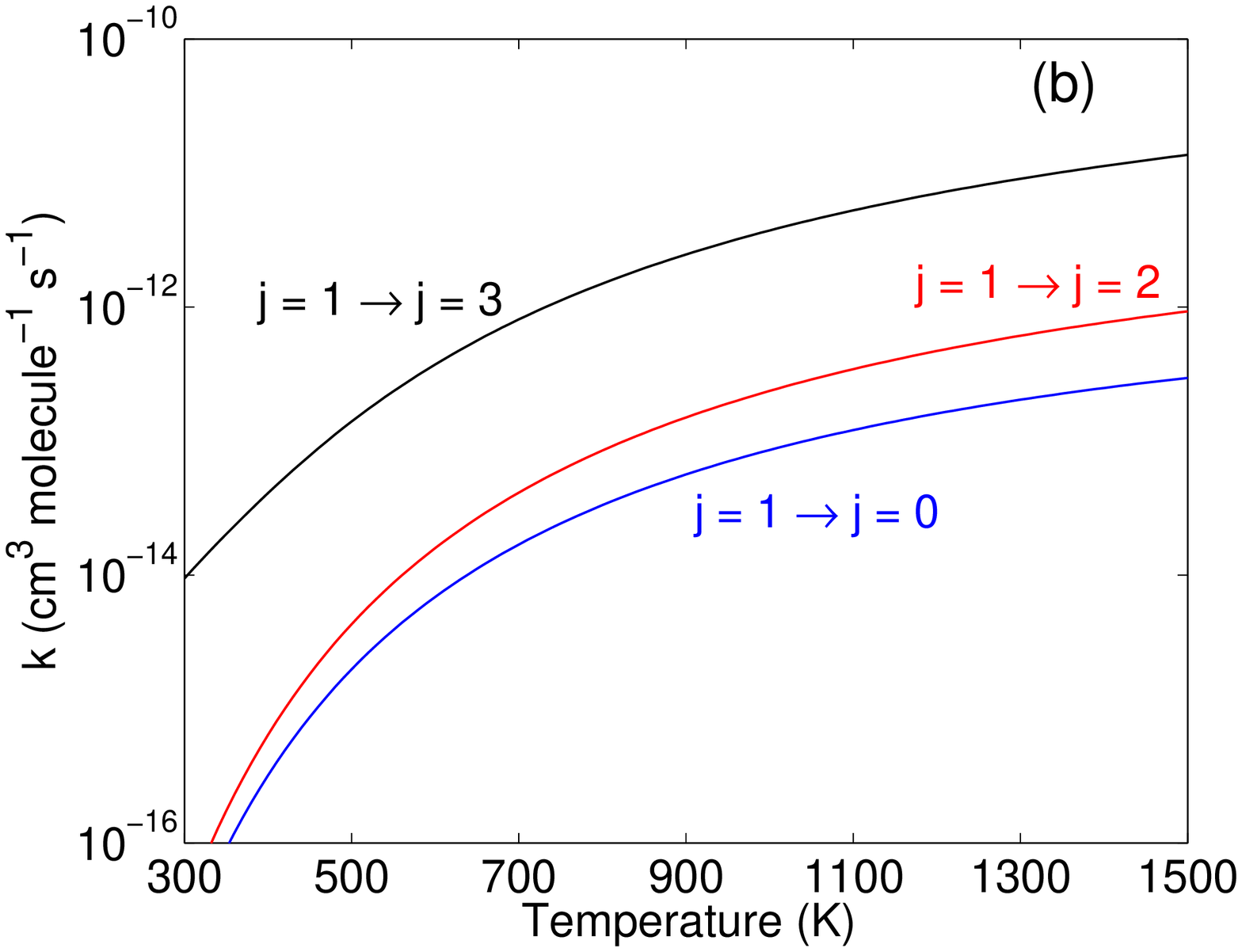}
\caption{Temperature dependence of the rate coefficients for the
  rotational excitation of p-H$_2$ (a) and o-H$_2$
 (b) by H. Reprinted with permission from Ref.  \cite{lique:12:H2}. Copyright 2012 American Institute of Physics. }
\label{fig3}
\end{figure}

Figure~\ref{fig3} presents the temperature dependence of the rate
coefficients corresponding to the cross sections presented in
Fig.~\ref{fig2}. As expected, the OPC processes are not negligible above $\sim$ 300~K
and will start to play a role in the thermalization of the OPR of
H$_2$ in media with temperature greater than 300~K. State to state H+H$_2$ rate constants between the first eleven
levels of H$_2(j=0-10)$ are given in Tables \ref{tab:taux1} and
\ref{tab:taux2} for temperatures equal to 10, 20, 30, 50, 100, 200,
300, 500, 700, 1000 and 1500 K.

\begingroup
\squeezetable
\begin{table}
\caption{State to state H$_2$ + H rate constants. The rates are in units of~cm$^{3}$
  s$^{-1}$.}
\label{tab:taux1}
 \begin{ruledtabular}
\begin{tabular}{cc|ccccccccccc}
$j$ & $j'$ & 10K & 20K & 30K & 50K & 100K & 200K & 300K & 500K & 700K & 1000 & 1500K \\
\hline
 1  & 0  & 1.24(-24) & 2.10(-24)  & 3.29(-24)  & 8.98(-24)  & 2.17(-22)  & 1.56(-19)  & 1.82(-17)  & 1.98(-15)  & 1.69(-14)  & 8.57(-14)  & 2.96(-13) \\
 2  & 0  & 6.02(-14) & 6.37(-14)  & 6.37(-14)  & 6.44(-14)  & 7.02(-14)  & 8.01(-14)  & 9.16(-14)  & 2.19(-13)  & 6.59(-13)  & 2.00(-12)  & 5.35(-12) \\
 2  & 1  & 8.21(-23)  & 9.67(-23)  & 1.18(-22)  & 2.37(-22)  & 3.99(-21)  & 2.07(-18)  & 1.82(-16)  & 1.52(-14)  & 1.20(-13)  & 5.98(-13)  & 2.10(-12) \\
 3  & 0  & 2.33(-22)  & 3.13(-22)  & 4.02(-22)  & 7.25(-22)  & 5.88(-21)  & 1.04(-18)  & 5.01(-17)  & 2.46(-15)  & 1.59(-14)  & 7.02(-14)  & 2.30(-13) \\
 3  & 1  & 8.11(-15)  & 1.06(-14)  & 1.26(-14)  & 1.62(-14)  & 2.49(-14)  & 4.04(-14)  & 6.75(-14)  & 3.25(-13)  & 1.16(-12)  & 3.72(-12)  & 1.03(-11) \\
 3  & 2  & 7.19(-23)  & 1.02(-22)  & 1.40(-22)  & 3.03(-22)  & 4.35(-21)  & 1.60(-18)  & 9.76(-17)  & 6.20(-15)  & 4.69(-14)  & 2.37(-13)  & 8.76(-13) \\
 4  & 0  & 1.29(-17)  & 1.81(-17)  & 2.31(-17)  & 3.58(-17)  & 1.01(-16)  & 6.03(-16)  & 2.89(-15)  & 3.12(-14)  & 1.26(-13)  & 4.42(-13)  & 1.27(-12) \\
 4  & 1  & 1.08(-20)  & 1.48(-20)  & 1.99(-20)  & 3.64(-20)  & 2.44(-19)  & 2.34(-17)  & 6.39(-16)  & 2.04(-14)  & 1.19(-13)  & 5.14(-13)  & 1.72(-12) \\
 4  & 2  & 1.74(-16)  & 2.95(-16)  & 4.40(-16)  & 8.41(-16)  & 2.83(-15)  & 1.59(-14)  & 6.05(-14)  & 4.07(-13)  & 1.35(-12)  & 4.01(-12)  & 1.06(-11) \\
 4  & 3  & 4.86(-22)  & 7.71(-22)  & 1.13(-21)  & 2.54(-21)  & 3.31(-20)  & 8.14(-18)  & 3.39(-16)  & 1.77(-14)  & 1.33(-13)  & 7.01(-13)  & 2.76(-12) \\
 5  & 0  & 2.09(-20)  & 2.94(-20)  & 3.81(-20)  & 6.43(-20)  & 3.26(-19)  & 1.26(-17)  & 1.61(-16)  & 2.88(-15)  & 1.43(-14)  & 5.73(-14)  & 1.86(-13) \\
 5  & 1  & 3.27(-17)  & 4.45(-17)  & 5.49(-17)  & 7.92(-17)  & 1.93(-16)  & 1.18(-15)  & 6.69(-15)  & 7.91(-14)  & 3.41(-13)  & 1.22(-12)  & 3.67(-12) \\
 5  & 2  & 2.34(-20)  & 3.35(-20)  & 4.43(-20)  & 7.83(-20)  & 4.61(-19)  & 2.39(-17)  & 3.88(-16)  & 9.23(-15)  & 5.18(-14)  & 2.25(-13)  & 7.82(-13) \\
 5  & 3  & 1.26(-15)  & 1.73(-15)  & 2.17(-15)  & 3.20(-15)  & 7.75(-15)  & 3.36(-14)  & 1.06(-13)  & 5.66(-13)  & 1.72(-12)  & 4.99(-12)  & 1.33(-11) \\
 5  & 4  & 3.90(-22)  & 6.05(-22)  & 8.71(-22)  & 1.89(-21)  & 2.17(-20)  & 3.39(-18)  & 1.07(-16)  & 5.03(-15)  & 3.86(-14)  & 2.13(-13)  & 8.97(-13) \\
 6  & 0  & 1.38(-17)  & 1.17(-17)  & 9.85(-18)  & 8.66(-18)  & 1.70(-17)  & 1.87(-16)  & 1.30(-15)  & 1.52(-14)  & 6.26(-14)  & 2.12(-13)  & 6.02(-13) \\
 6  & 1  & 7.27(-18)  & 6.24(-18)  & 5.32(-18)  & 4.98(-18)  & 1.31(-17)  & 2.06(-16)  & 1.63(-15)  & 2.34(-14)  & 1.12(-13)  & 4.43(-13)  & 1.45(-12) \\
 6  & 2  & 2.90(-16)  & 2.49(-16)  & 2.11(-16)  & 1.85(-16)  & 2.98(-16)  & 1.67(-15)  & 8.16(-15)  & 7.94(-14)  & 3.19(-13)  & 1.10(-12)  & 3.22(-12) \\
 6  & 3  & 1.83(-18)  & 1.59(-18)  & 1.39(-18)  & 1.40(-18)  & 4.71(-18)  & 1.24(-16)  & 1.47(-15)  & 3.01(-14)  & 1.68(-13)  & 7.45(-13)  & 2.68(-12) \\
 6  & 4  & 1.81(-14)  & 1.55(-14)  & 1.32(-14)  & 1.14(-14)  & 1.60(-14)  & 5.23(-14)  & 1.46(-13)  & 6.62(-13)  & 1.82(-12)  & 4.83(-12)  & 1.20(-11) \\
 6  & 5  & 1.13(-20)  & 1.01(-20)  & 9.39(-21)  & 1.18(-20)  & 9.17(-20)  & 1.01(-17)  & 2.78(-16)  & 1.26(-14)  & 9.93(-14)  & 5.75(-13)  & 2.55(-12) \\
 7  & 0  & 7.12(-19)  & 9.71(-19)  & 1.20(-18)  & 1.79(-18)  & 4.98(-18)  & 3.95(-17)  & 2.42(-16)  & 2.87(-15)  & 1.26(-14)  & 4.71(-14)  & 1.50(-13) \\
 7  & 1  & 1.17(-17)  & 1.62(-17)  & 2.03(-17)  & 3.07(-17)  & 9.04(-17)  & 7.73(-16)  & 4.61(-15)  & 5.13(-14)  & 2.14(-13)  & 7.54(-13)  & 2.25(-12) \\
 7  & 2  & 1.42(-18)  & 1.97(-18)  & 2.47(-18)  & 3.79(-18)  & 1.20(-17)  & 1.22(-16)  & 8.24(-16)  & 1.11(-14)  & 5.25(-14)  & 2.08(-13)  & 6.90(-13) \\
 7  & 3  & 5.50(-17)  & 7.55(-17)  & 9.37(-17)  & 1.37(-16)  & 3.46(-16)  & 2.18(-15)  & 1.09(-14)  & 1.09(-13)  & 4.58(-13)  & 1.67(-12)  & 5.21(-12) \\
 7  & 4  & 2.07(-19)  & 2.95(-19)  & 3.84(-19)  & 6.40(-19)  & 2.71(-18)  & 5.20(-17)  & 5.23(-16)  & 1.01(-14)  & 5.70(-14)  & 2.59(-13)  & 9.60(-13) \\
 7  & 5  & 3.27(-15)  & 4.47(-15)  & 5.49(-15)  & 7.72(-15)  & 1.64(-14)  & 5.91(-14)  & 1.64(-13)  & 7.31(-13)  & 2.01(-12)  & 5.45(-12)  & 1.41(-11) \\
 7  & 6  & 8.63(-22)  & 1.31(-21)  & 1.85(-21)  & 3.79(-21)  & 3.46(-20)  & 3.18(-18)  & 8.09(-17)  & 3.60(-15)  & 2.91(-14)  & 1.74(-13)  & 8.08(-13) \\
 8  & 0  & 4.99(-18)  & 6.23(-18)  & 7.27(-18)  & 1.00(-17)  & 2.65(-17)  & 2.01(-16)  & 1.09(-15)  & 1.07(-14)  & 4.16(-14)  & 1.33(-13)  & 3.67(-13) \\
 8  & 1  & 8.18(-18)  & 1.03(-17)  & 1.21(-17)  & 1.70(-17)  & 4.61(-17)  & 3.71(-16)  & 2.14(-15)  & 2.35(-14)  & 9.99(-14)  & 3.68(-13)  & 1.17(-12) \\
 8  & 2  & 1.95(-17)  & 2.44(-17)  & 2.85(-17)  & 3.96(-17)  & 1.05(-16)  & 7.92(-16)  & 4.40(-15)  & 4.54(-14)  & 1.82(-13)  & 6.18(-13)  & 1.78(-12) \\
 8  & 3  & 7.69(-18)  & 9.82(-18)  & 1.17(-17)  & 1.69(-17)  & 4.95(-17)  & 4.55(-16)  & 3.02(-15)  & 4.06(-14)  & 1.91(-13)  & 7.52(-13)  & 2.51(-12) \\
 8  & 4  & 7.28(-17)  & 9.27(-17)  & 1.10(-16)  & 1.53(-16)  & 3.77(-16)  & 2.25(-15)  & 1.03(-14)  & 9.03(-14)  & 3.51(-13)  & 1.20(-12)  & 3.57(-12) \\
 8  & 5  & 1.13(-18)  & 1.47(-18)  & 1.80(-18)  & 2.80(-18)  & 1.07(-17)  & 1.73(-16)  & 1.62(-15)  & 3.08(-14)  & 1.75(-13)  & 8.03(-13)  & 3.01(-12) \\
 8  & 6  & 3.77(-15)  & 4.81(-15)  & 5.66(-15)  & 7.64(-15)  & 1.60(-14)  & 5.82(-14)  & 1.64(-13)  & 7.26(-13)  & 1.94(-12)  & 5.00(-12)  & 1.22(-11) \\
 8  & 7  & 5.21(-21)  & 6.96(-21)  & 8.92(-21)  & 1.61(-20)  & 1.25(-19)  & 9.56(-18)  & 2.27(-16)  & 9.79(-15)  & 7.97(-14)  & 4.87(-13)  & 2.31(-12) \\
\end{tabular}
 \end{ruledtabular}
%\end{center}
\end{table}
\endgroup

 \begingroup
\squeezetable
\begin{table}
\caption{H$_2$ + H rate constants. The rates are in units of~cm$^{3}$
  s$^{-1}$.}
\label{tab:taux2}
 \begin{ruledtabular}
\begin{tabular}{cc|ccccccccccc}
$j$ & $j'$ & 10K & 20K & 30K & 50K & 100K & 200K & 300K & 500K & 700K & 1000 & 1500K \\
\hline
9  & 0  & 6.59(-19)  & 1.04(-18)  & 1.42(-18)  & 2.32(-18)  & 6.76(-18)  & 4.88(-17)  & 2.56(-16)  & 2.51(-15)  & 1.01(-14)  & 3.57(-14)  & 1.12(-13) \\
 9  & 1  & 1.06(-17)  & 1.65(-17)  & 2.23(-17)  & 3.63(-17)  & 1.07(-16)  & 8.07(-16)  & 4.30(-15)  & 4.17(-14)  & 1.62(-13)  & 5.38(-13)  & 1.55(-12) \\
 9  & 2  & 2.13(-18)  & 3.37(-18)  & 4.63(-18)  & 7.75(-18)  & 2.42(-17)  & 2.00(-16)  & 1.13(-15)  & 1.18(-14)  & 4.92(-14)  & 1.77(-13)  & 5.58(-13) \\
 9  & 3  & 1.51(-17)  & 2.36(-17)  & 3.22(-17)  & 5.33(-17)  & 1.63(-16)  & 1.32(-15)  & 7.52(-15)  & 7.99(-14)  & 3.29(-13)  & 1.16(-12)  & 3.50(-12) \\
 9  & 4  & 1.57(-18)  & 2.45(-18)  & 3.35(-18)  & 5.59(-18)  & 1.83(-17)  & 1.77(-16)  & 1.20(-15)  & 1.60(-14)  & 7.38(-14)  & 2.85(-13)  & 9.33(-13) \\
 9  & 5  & 4.63(-17)  & 7.15(-17)  & 9.63(-17)  & 1.55(-16)  & 4.34(-16)  & 2.78(-15)  & 1.34(-14)  & 1.26(-13)  & 5.14(-13)  & 1.83(-12)  & 5.65(-12) \\
 9  & 6  & 2.20(-19)  & 3.90(-19)  & 5.78(-19)  & 1.08(-18)  & 4.36(-18)  & 6.38(-17)  & 5.68(-16)  & 1.07(-14)  & 6.05(-14)  & 2.75(-13)  & 1.02(-12) \\
 9  & 7  & 1.86(-15)  & 2.80(-15)  & 3.69(-15)  & 5.67(-15)  & 1.33(-14)  & 5.24(-14)  & 1.53(-13)  & 7.08(-13)  & 1.95(-12)  & 5.20(-12)  & 1.32(-11) \\
 9  & 8  & 4.39(-21)  & 5.28(-21)  & 5.79(-21)  & 8.14(-21)  & 5.59(-20)  & 3.59(-18)  & 7.79(-17)  & 3.10(-15)  & 2.47(-14)  & 1.50(-13)  & 7.06(-13) \\
 10 & 0  & 4.82(-18)  & 6.29(-18)  & 7.43(-18)  & 1.02(-17)  & 2.66(-17)  & 1.97(-16)  & 1.02(-15)  & 9.02(-15)  & 3.19(-14)  & 9.55(-14)  & 2.45(-13) \\
 10 & 1  & 1.08(-17)  & 1.43(-17)  & 1.68(-17)  & 2.29(-17)  & 5.87(-17)  & 4.24(-16)  & 2.19(-15)  & 2.04(-14)  & 7.89(-14)  & 2.71(-13)  & 8.26(-13) \\
 10 & 2  & 1.99(-17)  & 2.57(-17)  & 3.01(-17)  & 4.05(-17)  & 1.03(-16)  & 7.51(-16)  & 3.93(-15)  & 3.65(-14)  & 1.35(-13)  & 4.27(-13)  & 1.16(-12) \\
 10 & 3  & 2.03(-17)  & 2.63(-17)  & 3.08(-17)  & 4.18(-17)  & 1.09(-16)  & 8.42(-16)  & 4.59(-15)  & 4.65(-14)  & 1.87(-13)  & 6.54(-13)  & 1.99(-12) \\
 10 & 4  & 2.25(-17)  & 2.95(-17)  & 3.52(-17)  & 4.89(-17)  & 1.31(-16)  & 9.99(-16)  & 5.37(-15)  & 5.29(-14)  & 2.08(-13)  & 7.02(-13)  & 2.02(-12) \\
 10 & 5  & 9.40(-18)  & 1.25(-17)  & 1.50(-17)  & 2.16(-17)  & 6.42(-17)  & 6.39(-16)  & 4.34(-15)  & 5.59(-14)  & 2.50(-13)  & 9.36(-13)  & 2.94(-12) \\
 10 & 6  & 9.28(-17)  & 1.21(-16)  & 1.43(-16)  & 1.95(-16)  & 4.79(-16)  & 2.98(-15)  & 1.33(-14)  & 1.08(-13)  & 3.92(-13)  & 1.26(-12)  & 3.54(-12) \\
 10 & 7  & 2.17(-18)  & 2.86(-18)  & 3.45(-18)  & 5.02(-18)  & 1.67(-17)  & 2.18(-16)  & 1.83(-15)  & 3.38(-14)  & 1.88(-13)  & 8.36(-13)  & 2.98(-12) \\
 10 & 8  & 2.42(-15)  & 3.19(-15)  & 3.80(-15)  & 5.14(-15)  & 1.12(-14)  & 4.53(-14)  & 1.36(-13)  & 6.34(-13)  & 1.69(-12)  & 4.24(-12)  & 9.88(-12) \\
 10 & 9  & 9.59(-21)  & 1.00(-20)  & 1.17(-20)  & 2.62(-20)  & 2.55(-19)  & 1.46(-17)  & 2.64(-16)  & 8.89(-15)  & 6.80(-14)  & 4.02(-13)  & 1.84(-12) 
\end{tabular}
 \end{ruledtabular}
%\end{center}
\end{table}
\endgroup

\begin{figure}
\includegraphics[width=8. cm]{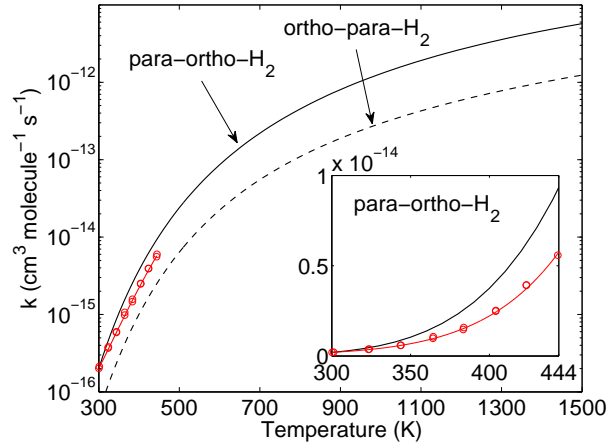}
\caption{Temperature dependence of the rate coefficients for the
  para--ortho-H$_2$ and ortho--para-H$_2$ conversion. The line with
  circles indicates the experimental results of Schulz and Le Roy
   \cite{Schulz:65}. Reprinted with permission from Ref. \cite{lique:12:H2}. {Copyright 2012  American Institute of Physics.}}
\label{fig4}
\end{figure}

The new theoretical results have then been compared with previous
experimental results. Figure \ref{fig4} present a
comparison of the theoretical results (properly averaged over
H$_2$ rotational distribution) (see Paper I) with the experimental
ones of Schulz \& Le Roy \cite{Schulz:65}.  The agreement with
experimental results \cite{Schulz:65} is a new nice illustration of
the detailed understanding of the simplest chemical reaction.

The new rotational rate constants were also compared with the
previously available data of Sun \& Dalgarno \cite{Sun:94} also
calculated using a quantum time-independent approach. The present data
differ from the previous results especially at low temperature (see
Paper I). The differences are a signature of the different H$_3$ PES
used in the two calculations. Sun \& Dalgarno \cite{Sun:94} used the
DMBE \cite{Varandas:87} PES whereas we used the most recent PES of
Mielke {\it et al.} \cite{Mielke02}. Then, we recommend the use of the
new rate constants for the astrophysical modeling.

Finally, it was interesting to compare the results of purely
rotationally inelastic scattering of H$_2$ by H using the rigid rotor
approximation and neglecting the reactive channels with the results of
Paper I. Indeed, as mentioned above, most of the work dealing with the
collisional excitation of H$_2$ by H has been performed using these
approximation. In such approximation, the OPC process of H$_2$
is neglected.

Then, we have performed calculations for pure rotational excitation of
H$_2$ by H using the rigid rotor approximation. We have chosen for the
H$_2$ internuclear separation $r_{H_2}= 1.449$~a$_0$, the ground state
vibrationally averaged value. The standard time-independent coupled
scattering equations were solved using the MOLSCAT code
\citep{molscat:94}.

\begin{figure}
\includegraphics[width=7. cm]{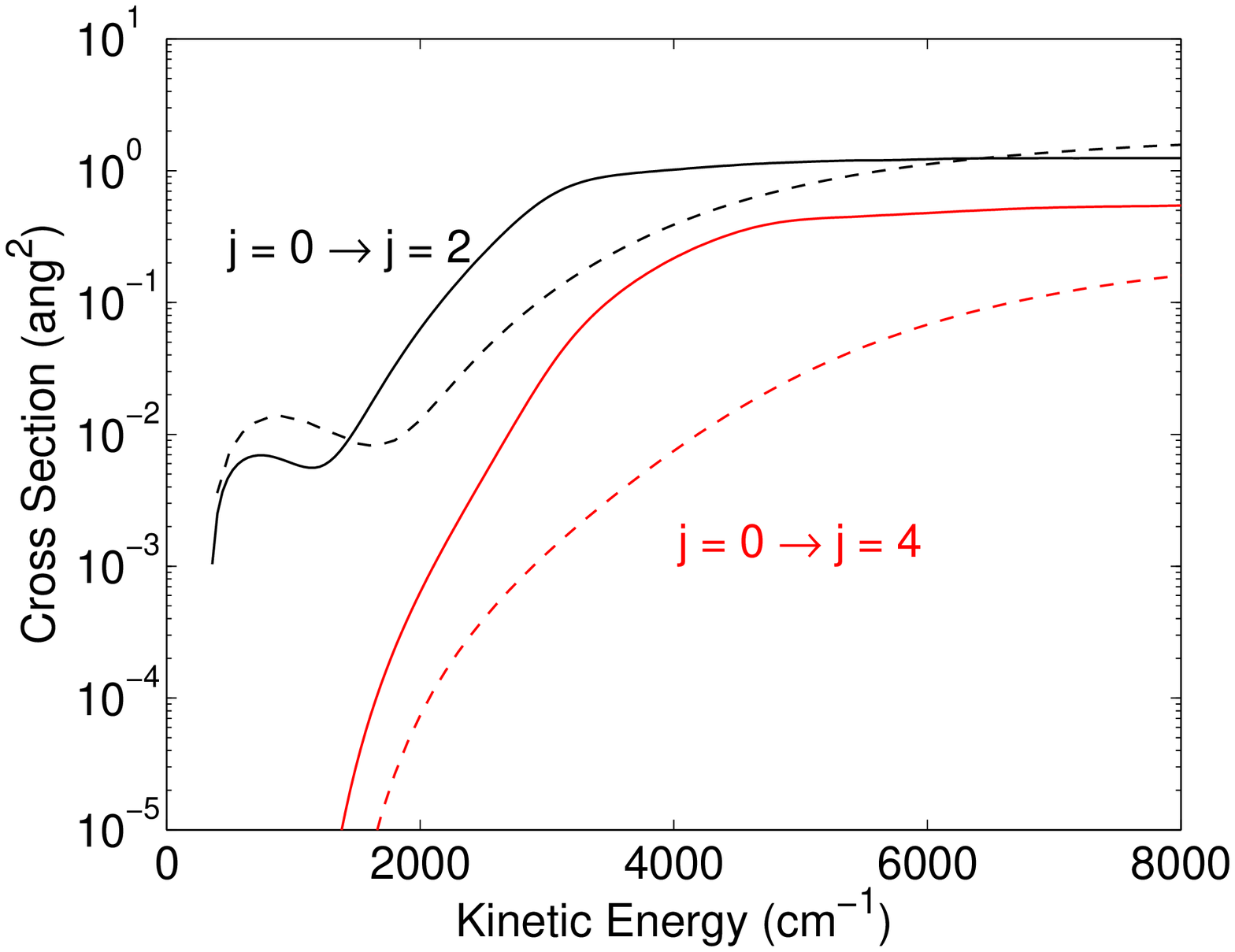}
\includegraphics[width=7. cm]{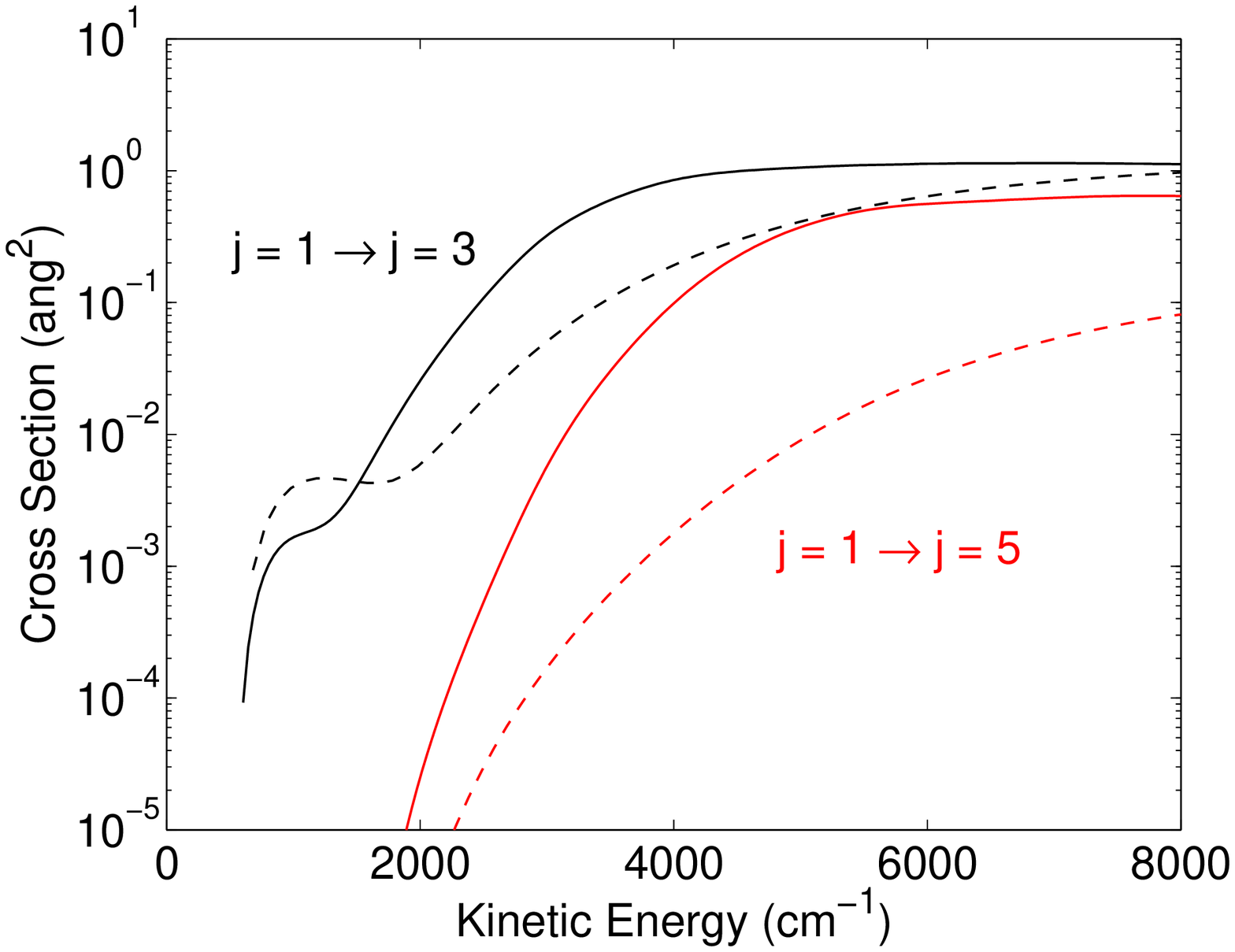}
\includegraphics[width=7. cm]{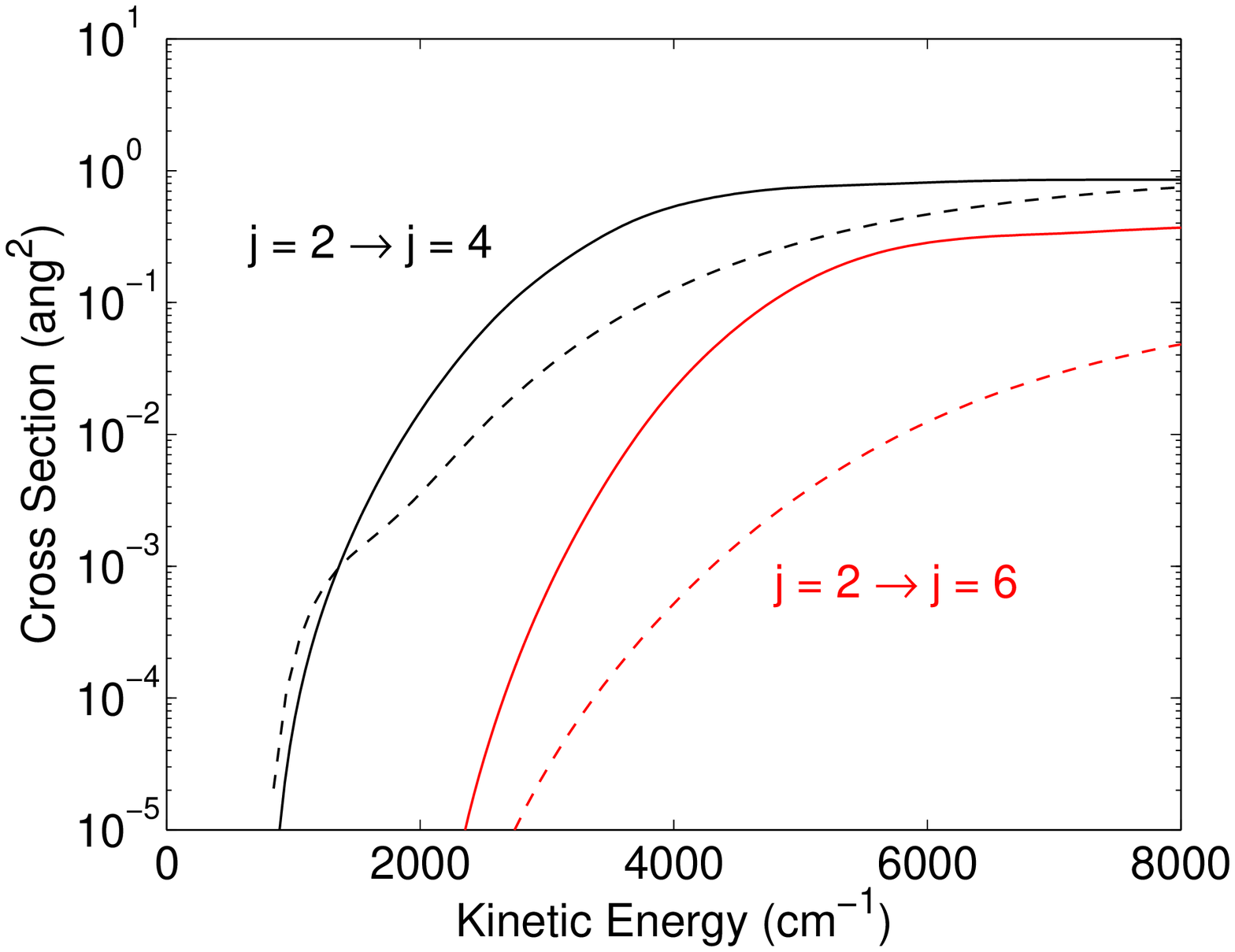}
\includegraphics[width=7. cm]{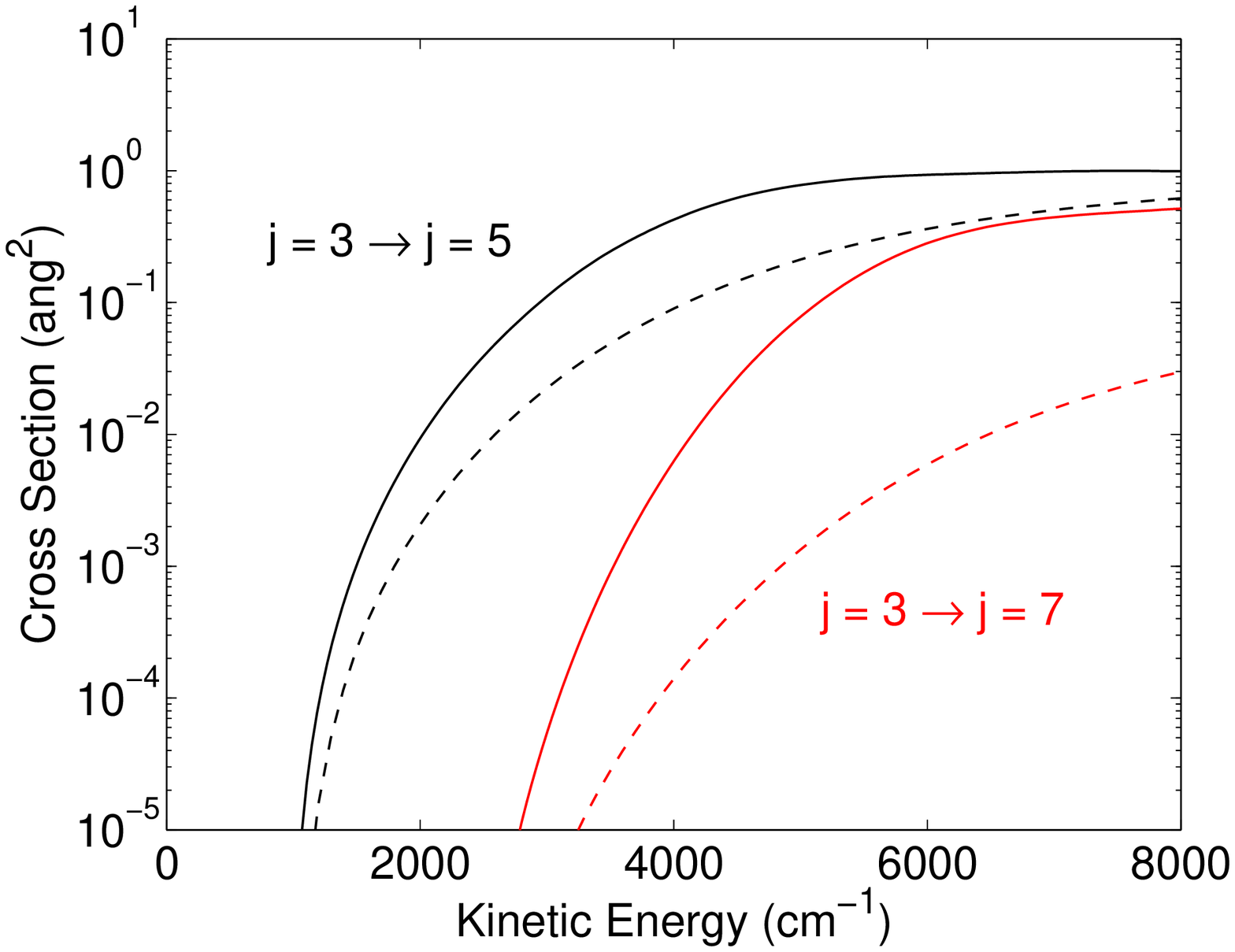}
\caption{Collision energy dependence of the integral cross section for the rotational excitation of para-H$_2(j=0,$ 2) and ortho-H$_2(j=1,$ 3) by H. The solid lines correspond to exact 3D results. The dashed lines correspond to 2D results.}
\label{fig5}
\end{figure}

Figures \ref{fig5} displays the energy dependence of the calculated
integral cross sections for rotational excitation obtained using a
rigid rotor approximation and obtained from a full 3D approach that
include the reactive channels.

%------------------------

One can see that the 2D approach leads to results in moderate
agreement with those of the 3D approach, the agreement being
  better for small $\Delta j$ transitions than for larger $\Delta j$
  transitions. This reflects the fact that transitions with
  large $\Delta j$, which have the smallest cross sections, are the
  most sensitive to modest changes in the PES.

At low collisional energies, just above the thresholds, the 2D and 3D
results are in reasonable agreement (especially for $\Delta j=2$
transitions).  This could have been anticipated since the reactive
process, that could disturb the pure rotational excitation, is
inhibited by the reaction barrier.

However, both sets of results rapidly disagree with increasing
  energy despite the collisional energy is still largely lower the
  barrier height.

Two obvious reasons can explain these differences : \\
(i) The 2D approach does not take into account the vibration of
H$_2$ in the scattering calculations.\\
(ii) The 2D approach neglects the reactive channels.

At these intermediate collisional energies (where the
  reactants have not enough energy to overcome the barrier), one could
  have expected that the 2D approach, which entirely neglects the
  reactive OPC process of H$_2$, would overestimate the purely
  inelastic process. However, the opposite behaviour is found. As a
  result, the deviation between the two sets of cross sections cannot
  be explained by the presence of the (closed) reactive channels. Such
  behaviour was already observed for the rotational excitation of
  D$_2$ by H also performed using rigid rotor approximation and full
  3D approach \cite{lique:12}.

In fact, the behaviour with collision energy and the
magnitudes of the 2D and 3D inelastic cross sections can be
related to the radial dependence of the (intermolecular)
potential expansion coefficients $v_i(R)$. Wrathmall \& Flower
\cite{Wrathmall06} found in their study on the rotational excitation
of H$_2$ by H, using the PES of Mielke {\it et al.}  \cite{Mielke02},
that the cross section for the $j=0 \to j'=2$ transition significantly
decreases at moderate energies due to the shape of the $v_2(R)$
expansion coefficient. The present cross sections follow exactly the
same behaviour. Wrathmall \& Flower \cite{Wrathmall06} also found that
the minima in the cross sections were lowered and shifted to lower
energies when the expansion coefficients were averaged over the
vibrational ground state wave function (instead of fixing the
internuclear separation at its equilibrium value). Hence the
difference between the present 2D and 3D results at moderate energies
certainly reflects these intramolecular effects and shows the
importance of using 3D potential energy surfaces.

At high kinetic energies (above $\sim$5000~cm$^{-1}$), the
  2D cross sections tend to converge towards the 3D cross
  sections. However, at even higher kinetic energies, the 2D results
  can exceed the actual (3D) cross sections, as observed in Fig.~5 for
  the $j=0 \rightarrow 2$ transition above $\sim 6000$~cm$^{-1}$. This
  behaviour is expected in this energy regime since the scattering
  flux is directed into both inelastic and reactive channels. On the
  other hand, as observed in Fig.~2, the reactive cross sections
  ($\Delta j=1$) are significantly smaller than the purely inelastic
  ones. This explains why the 2D and 3D results are in rather good
  agreement in the high energy regime, in spite of the opening of the
  reactive channels. In summary, the major 3D effects occur at
  intermediate energies (in the range $\sim 2000-6000$~cm$^{-1}$) and
  are caused by the intramolecular dependence of the PES expansion
  coefficients.

We conclude that the calculations of rotationally inelastic rate
coefficients using the 2D approach provide the correct order
  of magnitude for the dominant transitions ($\Delta j=2$) but the 3D
  approach is necessary for an accuracy at the state-to-state level
  better than a factor of $\sim$ 3. We also recommend the use of our
  new data for modeling the pure rotational excitation process of
  H$_2$ by H.

%-------------------

\subsection{Ortho--para-H$_2$ conversion by proton exchange}

%Among the possible OPC oprocesses of H$_2$ that can
%occur at low temperatures, radiative transitions between the {\it
%  ortho} and {\it para} forms of H$_2$ are known to be very slow, with
%an interconversion lifetime greater than the age of the Universe for
%the $j=1\rightarrow 0$ transition \cite{pachucki08}.

The reaction with H, presented just above, has a substantial
activation energy ($\sim$5000~K) and it is thus inefficient at low
temperatures as those found for instance in the interstellar dense
clouds (about 10 K). The same happens for reaction with H$_2$. As a
result, collisions of H$_2$ with protons, H$^+$, are expected to drive
the OPR of H$_2$ in most cold astrophysical environments, from the
primordial to the interstellar gas.  The knowledge of the rate
coefficients for the H$^+$+H$_2$ reaction, especially for the $j=1$
$\rightarrow$ $j'=0$ transition, is thus of fundamental interest,
and especially for the cold astrophysical media. Given the very low
population of the rotationally excited H$_2$ below 100 K, we consider
here only the $j=0$ and $j=1$ rotational states of H$_2$.

Like the H+H$_2$ reaction, the H$^+$+H$_2$ reaction is one of the
simplest elementary reaction and, surprisingly, there exist only
approximate calculations for the OPC process of H$_2$. The current
astrophysical models use values of the rate coefficients that have
been computed 20 years ago by Gerlich \cite{gerlich90} using a
statistical approach. Recently, we have computed for the first time
the rate coefficients for the OPC of H$_2$ by proton exchange at low
temperature ($T<100$~K) with a high
accuracy \cite{honvault:11,honvault12}. The fully time independent
quantum mechanical (TIQM) method combined with the most recent global
{\it ab initio} PES, both being described above, was
employed. We therefore used the same PES and the same TIQM method 
than for the study of the D$^+$+H$_2$ for which an excellent agreement 
has been obtained between the theoretical results and the measurements \cite{honvault13b}.
Statistical quantum mechanical (SQM)
\cite{RHM:CPL01,RGM:JCP03} calculations were also performed for the H$^+$+H$_2$ reaction and its isotopic variants
\cite{GRHLBAB:JCP06,ASGM:JCP07,honvault:11,honvault11a,honvault12}.
The SQM approach is based on the assumption that the process takes
place {\it via} a complex-forming mechanism and that seems to be
justified in the H$^+$+H$_2$ reaction. A comparison between the SQM
and the TIQM results could confirm or not the statistical behaviour of
H$^+$+H$_2$. The considerations mentioned above for symmetries and
nuclear spin also apply to the SQM method which is also based on a
quantum mechanical formalism.

We were interested in collision energies up to 0.1 eV, and thus the
charge transfer channel leading to H$_2^+$+H is not accessible (open
at 1.8 eV). However, we have to mention that non-adiabatic interaction
with the charge transfer channel could modify the energy level
spectrum of H$_3^+$ and it might influence the dynamics.  
However, as for the H + H$_2$ reaction, we expect that these
non-adiabatic process may not be very important, justifying the
Born-Oppenheimer approximation that was used.
In addition,
considering the temperature range chosen in this study ($T <$ 100 K),
only the fundamental vibrational quantum number $v=0$ of H$_2$ is considered.

\begin{figure}
\includegraphics*[width=80mm]{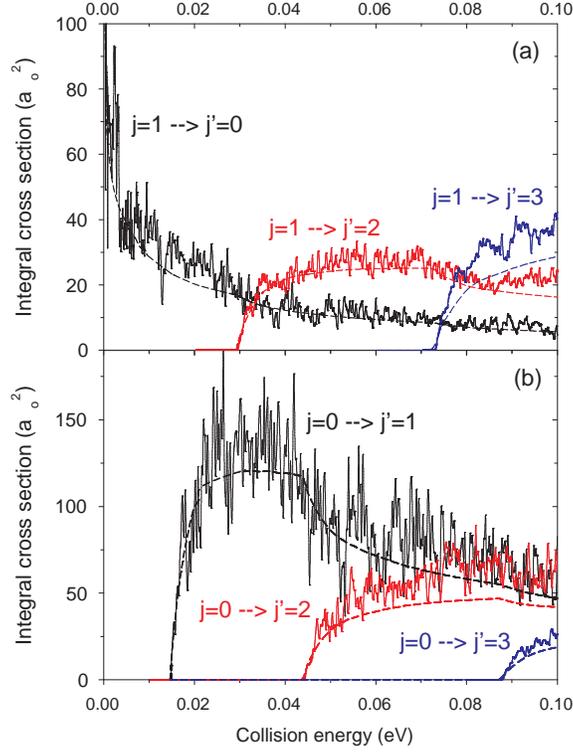}
\caption{TIQM (solid line) and SQM (dashed line) integral cross sections as a function of the collision energy for H$^+$+H$_2(v=0,j=1)$ $\rightarrow$ H$^+$+H$_2(v=0,j')$ (a) and H$^+$+H$_2(v=0,j=0)$ $\rightarrow$ H$^+$+H$_2(v=0,j')$ (b). Reprinted with permission from Ref. \cite{honvault:11}. Copyright 2011 American Physical Society.}
\label{fig1hp}
\end{figure}

Fig.~\ref{fig1hp}a shows the TIQM cross sections for the H$^+$+H$_2(v=0,j=1)$
$\rightarrow$ H$^+$+H$_2(v=0,j')$ reaction.  As expected for a
barrierless entrance channel, the cross sections for the $j=1 \rightarrow j'=0$
OPC process of H$_2$ decreases relatively smoothly with the collision
energy. There is indeed no energy threshold, in contrast with the case
of the $j=1 \rightarrow j'=2$ and $j=1 \rightarrow j'=3$ transitions
where energy thresholds of 0.029 eV and 0.073 eV respectively
exist. These values correspond to the energy difference between the
rotational levels involved in these transitions.  The main result
shown in Fig.~\ref{fig1hp}a is that the OPC of H$_2$, $j=1 \rightarrow
j'=0$, is the only possible process at the lowest collision energies.
Another interesting feature is the resonance structure found in the
cross sections. These resonances, which have survived to the partial wave $J$
summation, average the more narrow peaks observed in the reaction
probabilities (not shown here), which are linked to the presence of a
long-lived intermediate complex, H$_3^+$, formed in the deep well (4.6
eV) of the PES which supports many quasi-bound states.  
A recent experimental study \cite{Gerlich:13} has shown that the observed resonances here can be 
directly related with H$^+$--H$_2$ radiative association processes.

The situation is totally different for the H + H$_2$ reaction where no resonance
structure was found (see for instance Fig \ref{fig2}) because of the
existence of a very short-lived H$_3$ complex during the collision.
The SQM method by nature cannot reproduce the numerous
peaks. However, on average, a fairly good agreement is obtained in average for all
processes.

TIQM cross sections for the H$^+$+H$_2(v=0,j=0)$ $\rightarrow$ H$^+$+H$_2(v=0,j')$
 reaction are presented in Fig.~\ref{fig1hp}b.  In
contrast with the previous processes, an energy threshold is observed
here for all transitions because the reaction is always
endothermic. Many resonances are found again and the SQM prediction is
in good agreement with the TIQM results.  By combining the agreement
found also for the $j=1$ $\rightarrow$ $j'$, it is clear that the
H$^+$+H$_2$ reaction is statistical at low collision energies.  In
conclusion, the only possible transition at very low collision
energies (below 0.015 eV) is the $j=1 \rightarrow j'=0$ OPC process of H$_2$ ,
while at higher collision energies the $j=0 \rightarrow j'=1$ process
is dominant.

\begin{figure}
\includegraphics*[width=80mm]{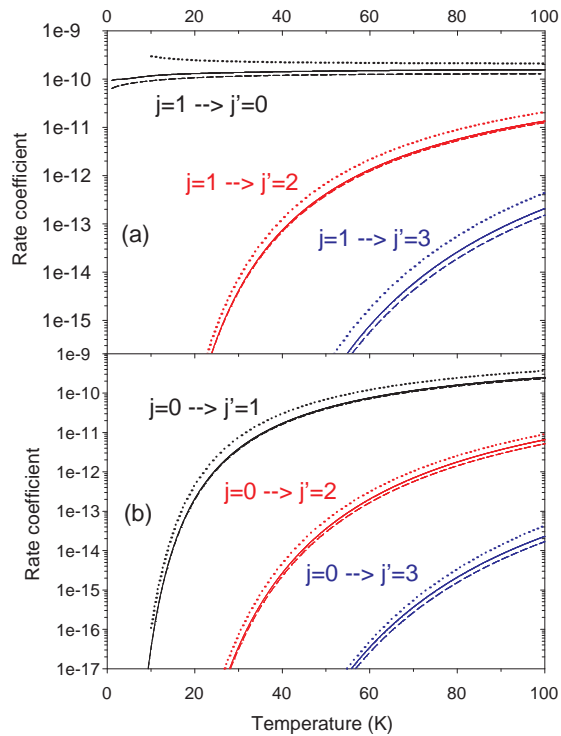}
\caption{TIQM (solid line), SQM (dashed line) and SM (dotted line) rate coefficients (in cm$^3$.molecule$^{-1}$.s$^{-1}$) as a function of the temperature for H$^+$+H$_2(v=0,j=1)$ $\rightarrow$ H$^+$+H$_2(v=0,j')$ (a) and H$^+$+H$_2(v=0,j=0)$ $\rightarrow$ H$^+$+H$_2(v=0,j')$ (b). Reprinted with permission from Ref. \cite{honvault12}. Copyright 2012 American Physical Society.}
\label{fig2hp}
\end{figure}

Fig.~\ref{fig2hp}a displays the TIQM rate coefficients calculated from
the cross sections for the H$^+$+H$_2(v=0,j=1)$ $\rightarrow$
H$^+$+H$_2(v=0,j')$ reaction for temperatures up to 100 K. These rate
coefficients have been obtained by integration of the cross sections
shown in Fig.~\ref{fig1hp}a using a Maxwell-Boltzmann distribution
over the collision energies.  As expected from the absence of energy
threshold in the cross sections for the H$^+$+H$_2(v=0,j=1)$
$\rightarrow$ H$^+$+H$_2(v=0,j'=0)$ reaction, the OPC process of
H$_2$, $j=1 \rightarrow j'=0$, is the dominant process at low
temperature. The rate coefficient is almost independent of temperature
with a constant value of about 1.5 10$^{-10}$
cm$^3$.molecule$^{-1}$.s$^{-1}$, which is about 1/17 of the
(temperature independent) Langevin rate value (2.5 10$^{-9}$
cm$^3$.molecule$^{-1}$.s$^{-1}$).  The two other rate coefficients for
the $j=1 \rightarrow j'=2$ and $j=1 \rightarrow j'=3$ transitions
increase steadily with temperature with changes of some order of
magnitudes. They follow the same order than that found for the cross
sections, and still remain much lower than the rate for $j=1
\rightarrow j'=0$.

Following the good agreement in average between the SQM  and TIQM
cross sections, the SQM rate coefficients are also in good agreement with the
TIQM prediction. This comparison and the previous ones allow to
unambiguously establish the statistical nature of the reaction at low
temperature. Nevertheless, it is worth mentioning nevertheless that this work
focusses entirely on state-to-state quantities, which, in principle,
constitute the most severe test for a statistical technique. The
agreement with the corresponding exact quantum mechanical rate
coefficients for $T > 10$ K is therefore a remarkable result. 

%In addition, this good accord shows that the use of an accurate global
%{\it ab initio} PES and the rigorous quantum mechanical formalism of
%the SQM method constitute a significant improvement with respect to
%the study performed by Gerlich \cite{gerlich90}.

The rate coefficients computed from another statistical model (SM)
developped by Gerlich \cite{gerlich90}, more simple than the SQM
method, are also shown. These rates follow the general trends of the
TIQM and SQM rate coefficients. However, some differences exist.
First, for the $j=1 \rightarrow j'=0$ OPC process of H$_2$, the dependence on
temperature for the SM rate coefficient at the lowest temperatures
(below 30 K) is reverse of that found by the SQM and TIQM
methods. Second, for all transtions and for the whole temperature
range considered here, the SM rate coefficient always overestimates
the SQM and TIQM rate coefficients.

Astrophysical models, including the {\it ortho}-{\it para} distinction
of H$_2$, may present significant effects relatively to these
differences found between the different computed rate coefficients,
especially for the $j=1 \rightarrow j'=0$ OPC process of H$_2$. We thus
encourage the astrophysicists to use in their future models the TIQM
rate coefficients which are as accurate as possible and constitute a
valid benchmark.

The TIQM rate coefficients for the H$^+$+H$_2(v=0,j=0)$
$\rightarrow$ H$^+$+H$_2(v=0,j')$ are displayed in
Fig.~\ref{fig2hp}b for temperatures up to 100 K. The main comments
mentioned above apply here again. The comparison between
Fig.~\ref{fig2hp}a and Fig.~\ref{fig2hp}b definitively confirms that
the OPC process of H$_2$, $j=1 \rightarrow j'=0$, is the main process for
temperatures below 50 K. Above this temperature, the $j=0 \rightarrow
j'=1$ para-ortho conversion process becomes also important.

Given their importance for interstellar models and for practical
purposes, fits of the TIQM rate coefficients between $T=10$ K and
$T=100$~K for the state-to-state H$^+ + $H$_2(v=0,j=0,1) \rightarrow$
H$^+ + $H$_2(v'=0,j')$ reactions to the analytical Kooij formula,
$k(T) =\alpha (T/300)^{\beta} exp(-\gamma/T)$, have been
performed. Values for the corresponding parameters are shown in Table
\ref{tabfit}.

\begin{table}[t]
\caption{Parameters for the fits of the TIQM rate coefficients to the Kooij analytical expression 
$k(T) =\alpha (T/300)^{\beta} exp(-\gamma/T)$ for the different  H$^+ + $H$_2(v=0,j) \rightarrow$H$^+ + $H$_2(v'=0,j')$ reactions (indicated as $j \rightarrow j'$ in the table).
$\alpha$ is measured in cm$^{3}$ molecule$^{-1}$ s$^{-1}$ and $\gamma$ 
in K. Parenthesis $(x)$ stand for 10$^{x}$.  
Temperature range at which the proposed fitting expressions yield errors below $\sim 5 \%$ 
are listed at the last line. Reprinted with permission from Ref. \cite{honvault11a}. Copyright 2011 Royal Society of Chemistry.} 
\label{tabfit}\tabcolsep=5pt
\begin{center}
\begin{tabular}{c|cccccc} 
\hline \hline
& 0 $\rightarrow$ 1 & 0 $\rightarrow$ 2 & 0$ \rightarrow$ 3 & 
1 $\rightarrow$ 0 & 1 $\rightarrow$ 2 & 1 $\rightarrow$ 3 \\ \hline
$\alpha$ & 1.490(-9) & 1.185(-9) &  2.526(-10) & 1.823(-10) & 0.434(-9)  & 0.706(-9)  \\
$\beta$ & 2(-4)  & -0.0168 & -1.3214 & 0.1289 & 0.0030 & -0.7877 \\
$\gamma$ & 178.25 & 522.0667 & 1075.3996 & -0.0214 & 346.8679 & 899.4216 \\
T range (K) & 20-100 & 30-100 & 40-100 & 10-100 & 10-100 & 50-100  
 \\
\hline \hline
\end{tabular}
\end{center}
\end{table}

\section{Discussion and astrophysical applications}

As discussed in the introduction, molecular hydrogen is the most abundant
molecule in the Universe and its OPR plays a fundamental role in both
the physics and chemistry of the ISM and of the early universe. In this section we discuss the
relative role of the different OPC processes, that is {\it via}
reactive collisions with hydrogenated species (H, H$^+$, H$_2$,
H$_3^+$) and {\it via} interaction with solid surfaces. In the cold
ISM ($T\sim 10~K$), only H$^+$, H$_3^+$ and the icy dust grains can
contribute to the OPC of H$_2$ because of the large energy barriers to
hydrogen exchange in H+H$_2$ ($\sim 5000$~K \citep{lique:12:H2}) and
H$_2$+H$_2$ ($\sim$60 000~K \cite{carmona07}). In the gas phase,
conversion by protons and H$_3^+$ have similar rate coefficients of
$\sim 10^{-10}$~cm$^3$s$^{-1}$ for the $j=1\rightarrow 0$ ortho-para
transition in the temperature range 10$-$100~K (see Fig.~7 and
\cite{gomez12}). In typical cold molecular clouds, the H$_2$ density
is $\sim$10$^4$~cm$^{-3}$ and the relative abundances of H$^+$ and
H$_3^+$ (relative to H$_2$) are both a few 10$^{-9}$. These two ions
therefore contribute with similar OPC rates of $\sim
10^{-15}$~s$^{-1}$ and the timescale for the gas phase OPC in these
objects is $\sim 10$~Myr, which is close to the typical lifetime of
molecular clouds ($\sim 5$~Myr \cite{bergin07}). As a result, the OPR
of H$_2$ in the cold ISM is expected to be close to its steady-state
value, which should however differ from the thermal equilibrium value
\citep{flower06,faure13}.

In warmer regions such as protostellar shocks or PDRs, the kinetic
temperature can exceed 300~K and in this regime collisions with
hydrogen atoms become efficient with an OPC rate coefficient larger
than $\sim 10^{-16}$~cm$^3$s$^{-1}$ (see Fig.~4). If the abundance
ratio between H and H$^+$ (or H$_3^+$) is larger than $\sim 10^6$
(assuming a typical OPC rate coefficient of $\sim
10^{-10}$~cm$^3$s$^{-1}$ for protons above 100~K), hydrogen atoms will
then compete or even dominate the ions in the OPC of H$_2$ in these
environments. At temperatures above $\sim$600~K, the rate coefficient
for the conversion of p-H$_2$ to o-H$_2$ even exceeds
$10^{-13}$~cm$^3$s$^{-1}$ (Fig.~4). In media where the abundance ratio
[H] / [H$^+$] is larger than $\sim 1000$, hydrogen atoms will be thus
the major para-to-ortho converters with an OPC rate of $\sim
3200\times(100$~cm$^{-3}/n({\rm H}))$~yr, where $n({\rm H})$ is the density
of hydrogen atoms. Thus, if $n({\rm H})$ is larger than $\sim
10$~cm$^{-3}$, the para-to-ortho conversion by H can yield an
equilibrium OPR of 3 within less than 30 000~yr (at kinetic
temperatures above 600~K). We note that we do not consider here
H$_2$+H$_2$ collisions because these have a negligible OPC rate
coefficient of $\sim 10^{-27}$~cm$^3$s$^{-1}$ at 300~K
\citep{huestis08}. These collisions can however play a role in the
atmospheres of giant (exo)planets owing to H$_2$ densities much higher
than in the ISM.

If we now consider typical dust grains with size $\sim 0.1~\mu$m, a
dust-to-gas ratio of $\sim 10^{-12}$ by number and a H$_2$ mean
velocity of $\sim 3\times 10^4$~cm.s$^{-1}$, the typical collision
rate between H$_2$ and a dust grain is a few 10$^{-14}$~s$^{-1}$ at
10~K. Using a typical sticking probability of 0.1 \cite{fukutani13}
and a solid phase OPC efficiency of 1, the overall OPC rate is $\sim
10^{-15}$~s$^{-1}$, i.e. similar to the gas phase OPC rate in the cold
gas. Dust grains can thus possibly play a role in the OPC of H$_2$, in
competition with gas phase processes. It should be noted, however,
that the OPC efficiency on solid surfaces can be much lower than 1,
depending on the ratio between the residence time (which is highly
uncertain) and the conversion time (which is typically 10$^3$~s on
amorphous water and graphite) \cite{lebourlot00,fukutani13}. On the
other hand, nascent H$_2$ molecules formed on a grain could be
immediately retrapped in the very irregular surface of amorphous
solid water \cite{watanabe10}. The OPR of such retrapped H$_2$ could
then reflect the temperature of the surface. In summary, in contrast
to the OPC of H$_2$ by H and H$^+$ for which rate coefficients are now
known with high accuracy, the contribution of dust particles is
unclear and more experimental work on interstellar dust analogues is
urgently needed. Dedicated astrophysical models including all relevant processes, such as \citep{lebourlot00}, are also necessary to assess the impact of the new calculated rate coefficients.

Finally, in addition to ortho-para conversion, hydrogen atoms and
protons are important colliders than can also rotationally excite
H$_2$ molecules, in competition with other H$_2$ molecules and He
atoms (and electrons in highly ionized media). Thus, from data
reported in Tables~1-3, we can compare the rate coefficients for
ortho-ortho and para-para transitions ($\Delta j=2$) with those
computed for H$_2$+He and H$_2$+H$_2$ \citep{bala99,lee08}. Above
300~K, these latter are similar to those for H$_2$+H so that all the
dominant neutrals (H, He and H$_2$) significantly contribute to the
pure rotational excitation of H$_2$. At low temperature ($T\leq
100$~K) where data for proton exchange are available, the rate
coefficients for the transitions $j=0\rightarrow 2$ and
$j=1\rightarrow 3$ are 3-4 orders of magnitude larger for protons than
for neutrals. The contribution of protons in the rotational excitation
of H$_2$ ($\Delta j=2$) is therefore significant if their relative
abundance (with respect to the dominant neutral) exceeds $\sim
10^{-4}$, e.g. in the early universe \citep{galli13}.

\section{Conclusion}

We have attempted to review the recent achievements related to the gas
phase ortho--para conversion of H$_2$ in astrophysical media. Despite
these new data already improve significantly the modeling of this
process in interstellar media and early universe, there is still a
lack of data for high temperatures and for high ro-vibrational levels,
including for pure ro-vibrational excitation. Extrapolation to higher
(lower) temperatures is addressed for example in
Ref. \cite{schoier:05} where specific analytic extrapolation formulae
are suggested. The danger of extrapolation is however emphasized by
Flower and Pineau des For{\^e}ts \cite{Flower:12extr} for a number of
molecules including H$_2$ and we endorse these warnings on an even
larger scale.

Thus, it seems important to extend the present calculations in order
to provide the astronomers the necessary tools to model the
ro-vibrational excitation and ortho--para-H$_2$ conversion processes
in hot environments. Such extension should be feasible for the
ortho--para-H$_2$ conversion by hydrogen exchange since the
calculations are relatively fast in terms of CPU time. At the
opposite, for ortho--para-H$_2$ conversion by proton exchange, an
extension to higher temperature may be more difficult since it will
imply to deal with a huge number of coupled channels using the quantum
time independent approach. Statistical method may be then an
interesting alternative since quantum and statistical approach are generally in good agreement for this system.

\begin{acknowledgments}

FL, PH and AF acknowledge the CNRS national program ``Physique et
Chimie du Milieu Interstellaire''. F.L. is grateful for the financial
support of the CPER Haute-Normandie/CNRT/Energie, Electronique,
Mat\'eriaux. F.L. and A.F. acknowledge support by the Agence Nationale
de la Recherche (ANR-HYDRIDES), contract ANR-12-BS05-0011-01.

\end{acknowledgments}

%\bibliography{h3}% Produces the bibliography via BibTeX.

\end{document}